\documentclass[useAMS,usenatbib]{mn2e}
\usepackage{natbib}
\usepackage{graphicx}                   %for PS/EPS graphics inclusion, new
\usepackage{amsmath}
\usepackage{mathrsfs}
\usepackage{amssymb}
\usepackage{float}
\usepackage[multiple]{footmisc}

\usepackage{color}

\begin{document}

\title[Pulsar-black hole gravity tests]
{Pulsar-black hole binaries: prospects for new gravity tests with
future radio telescopes}

\author[Liu et al.]{
K.~Liu,$^{1,2}$
R.~P.~Eatough,$^{2}$
N.~Wex,$^2$ and
M.~Kramer,$^{2,3}$
\\
$^{1}$Station de radioastronomie de Nan\c{c}ay, Observatoire de
  Paris, CNRS/INSU, F-18330 Nan\c{c}ay, and Laboratoire de Physique et \\
  Chimie de l'Environnement et de l'Espace LPC2E CNRS-Universit\'{e} d'Orl\'{e}ans, F-45071 Orl\'{e}ans Cedex 02, France \\
$^{2}$Max-Planck-Institut f\"ur Radioastronomie, Auf dem H\"ugel 69,
      D-53121 Bonn, Germany \\
$^{3}$Jodrell Bank Centre for Astrophysics, The University of Manchester, Alan Turing Building, Manchester, M13 9PL, UK}

\maketitle

\begin{abstract}
The anticipated discovery of a pulsar in orbit with a black hole is
expected to provide a unique laboratory for black hole physics and
gravity. In this context, the next generation of radio telescopes,
like the Five-hundred-metre Aperture Spherical radio Telescope
(FAST) and the Square Kilometre Array (SKA), with their
unprecedented sensitivity, will play a key role. In this paper, we
investigate the capability of future radio telescopes to probe the
spacetime of a black hole and test gravity theories, by timing a
pulsar orbiting a stellar-mass-black-hole (SBH). Based on mock data
simulations, we show that a few years of timing observations of a
sufficiently compact pulsar-SBH (PSR-SBH) system with future radio
telescopes would allow precise measurements of the black hole mass
and spin. A measurement precision of one per cent can be expected
for the spin.  Measuring the quadrupole moment of the black hole,
needed to test GR's no-hair theorem, requires extreme system
configurations with compact orbits and a large SBH mass.
Additionally, we show that a PSR-SBH system can lead to greatly
improved constraints on alternative gravity theories even if they
predict black holes (practically) identical to GR's. This is
demonstrated for a specific class of scalar-tensor theories.
Finally, we investigate the requirements for searching for PSR-SBH
systems. It is shown that the high sensitivity of the next
generation of radio telescopes is key for discovering compact
PSR-SBH systems, as it will allow for sufficiently short survey
integration times.
\end{abstract}

\begin{keywords}
pulsars: general --- stars: black holes --- gravitation
\end{keywords}

%%%%%%%%%%%%%%%%%%%%%%%%%%%%%%%%%%%%%%%%%%%%%%%%%%%%%%%%%%%%%%%%%%%%%%%%%%%%%%%%

\section{Introduction}
\label{sec:intro}

The discovery of the first binary pulsar in 1974 marked the
beginning of a completely new field in experimental gravity
\citep{ht75a}. The clock-like nature of pulsars makes relativistic
binary pulsar systems ideal test beds of relativistic theories of
gravity, i.e.~GR and its alternatives \citep{dam09}. A system of a
pulsar in orbit with a black hole will be a unique celestial
laboratory for such tests, by both allowing for qualitatively new
tests of gravity and greatly improving the previous results
\citep{kbc+04}. \cite{wk99} first pointed out that pulsar timing can
be used to measure the properties of the companion black hole, and
test GR's cosmic censorship conjecture and no-hair theorem. It also
has been shown that a pulsar-black hole system is a great tool in
constraining alternative gravity theories, such as scalar-tensor
theories of gravity \citep{de98,esp09} and extra spatial dimensions
\citep{skm+11}.

Pulsars are usually weak radio sources and the precision of most
current tests are still limited by instrumental sensitivity.
Therefore, the next generation of radio telescopes, which will
provide a significant increase in source flux gain, are ideal tools
to carry out the aforementioned experiments. Theoretically, they
will provide one to two orders of magnitude improvement in timing
precision for millisecond pulsars (MSPs) of normal flux density
\citep{lvk+11}. The SKA and the FAST are the two representatives
which this paper will mainly refer to. It is certain that the next
generation of radio telescopes will open a new era on gravity theory
studies, mainly driven by the new pulsar timing and searching
capabilities of these future instruments \citep{kbc+04,laz13}.

There are mainly two scenarios where a pulsar is in orbit with a
black hole: a pulsar and stellar-mass black hole binary, and a
pulsar moving around a significantly more massive black
hole at the center of either the Galaxy or a
globular cluster. Concerning the second scenario, timing experiments
and gravity tests with a pulsar in orbit about the supermassive
black hole Sgr~A$^\ast$ in the Galactic centre have already been
discussed in detail by \cite{lwk+12}. They show that a few years of
timing of a pulsar near Sgr~A$^\ast$ (orbital period of a few
months) with the SKA should allow for the determination of the black
hole mass and spin with high precision (fractional precision of
$\sim 10^{-5}$ and $\sim 10^{-3}$, respectively) and may lead to a
test of the no-hair theorem with $\lesssim 1\%$ precision. In this
paper, we will mainly focus on the first scenario, a system
consisting of a pulsar and a stellar-mass black hole (PSR-SBH
system).

An order of $\sim10^8$ isolated SBHs are estimated to exist in our
Galaxy \citep{st83,vdh92}. So far 24 SBHs, as members of X-ray
binaries, have been identified via their masses, most of them within
our own Galaxy \citep{nm13}. The masses found range from about 5 to
30 solar masses. Based on models of the accretion flow, which
depends very strongly on how spin affects the strong-field black
hole spacetime, for several of the SBH candidates the spin parameter
$\chi$ has been determined. Quite a few of the SBH candidates seem
to rotate at a large fraction of the maximum spin ($\chi = 1$), with
$\chi
> 0.983\;(3{\sigma})$ for Cyg X-1 being the largest one \citep{gmr+11,gmr+13}.
So far, for none of the SBH candidates measurements are accurate
enough to allow for a test of the no-hair theorem. More details and
references for the current SBH candidates are provided by
\cite{nm13}.

There exist a few channels through which a PSR-SBH binary can be
formed. The first is to follow the standard binary evolutionary path
that has been widely studied by previous works \citep{yp98,vt03},
which may result in a PSR-SBH system of wide and eccentric orbit and
a young pulsar with slow spin-periods ($\sim 0.1$--$1$\,s). The
second approach is a so-called ``reversal mechanism'' where under a
certain set of circumstances the pulsar is formed first and later
spun up by accretion during the red giant phase of the other star
\citep{spn04,ppr05}. This may result in a system consisting of a
recycled pulsar in orbit with a black hole, which is more desirable
given that recycled pulsars are generally more precise timers than
slow ``normal'' pulsars \citep[e.g.][]{vbc+09}. Thirdly, a PSR-SBH
system can be formed through multiple body encounter, which is known
to occur in regions of high stellar density, such as globular
clusters and the Galactic centre region \citep{fl11,csc14}. As a
matter of fact, in many globular clusters we already know of several
cases, where the pulsar was first fully recycled to have millisecond
rotational period in a low-eccentricity binary system with a low
mass companion and later disrupted by the intrusion of a more
massive stellar remnant. This normally results in binary millisecond
pulsars (MSPs) with eccentric orbits and massive companions like
M15C \citep{pakw91}, NGC 1851A \citep{fgri04,frg07} and NGC 6544B
\citep{lfr+12}, and is exactly the same mechanism that is expected
to produce a MSP-SBH system in a globular cluster. In this paper, we
will not elaborate on the formation scenario or population synthesis
of PSR-SBH systems, but focus on the technical requirements and the
PSR-SBH system configurations that are necessary for different tests
of black hole physics and theories of gravity.

One of the major difficulties in finding PSR-SBH systems is the lack
of computational power \citep{eat09}. Nevertheless, recently
developed techniques such as the global volunteer distributed
computing project
Einstein@home\footnote{http://einstein.phys.uwm.edu/} have
already been involved in finding binary pulsars \citep{kac+10}. In
addition, the sensitivity of future telescopes will allow us to
overcome many of the computational limits imposed by long pointing
times through significantly shorter integrations, which will greatly
enhance our chance for the discovery of PSR-SBH systems.

The structure of the paper is as follows. In Section~\ref{sec:tels}
we introduce the design of the FAST and the SKA as two
representatives of the next generation of radio telescopes, as well
as their corresponding achievable timing precisions. In
Section~\ref{sec:GR_test} we discuss the measurability of black-hole
properties and tests of GR's cosmic censorship conjecture and the
no-hair theorem by timing a PSR-SBH system. Section~\ref{sec:stg}
shows the expected constrains on scalar-tensor theories of gravity
by a PSR-SBH system. A discussion on external effects that may
influence the aforementioned gravity tests is presented in
Section~\ref{sec:dis}. Section~\ref{sec:search} demonstrates that
the sensitivity allowed by the next generation of radio telescopes
would provide a significantly better chance to recover the pulsar
signal in a strong gravitational field. Our conclusion is shown in
Section~\ref{sec:conclu}.

%%%%%%%%%%%%%%%%%%%%%%%%%%%%%%%%%%%%%%%%%%%%%%%%%%%%%%%%%%%%%%%%%%%%%%%%%%%%%%%%

\section{Instrumentation and timing limits with future telescopes}
\label{sec:tels}

The next generation of radio telescopes will provide an increase in
collecting area by factor of 10 to 100, compared with the current
largest steerable single-dish antennas, and a factor of a few to 25
compared to Arecibo, the current largest single-aperture radio
telescope. Ideally, this will translate into the same orders of
magnitude improvement in pulsar-timing precision, which would
greatly boost the quality of current gravity tests and enable new
types of experiment \citep{kbc+04}.

%------------------------------------------------------------------------------%

\subsection{Telescope designs}
\label{ssec:NGRTs}

In a few years from now, the Five-hundred-metre Aperture Spherical
radio Telescope (FAST) will be the largest single dish radio
telescope on Earth \citep{nan06,nlj+11}. FAST is designed as an
Arecibo-type antenna and built into a karst depression in southern
China, which is sufficiently large to hold the 500-m diameter dish.
The effective aperture of FAST is equivalent to a fully illuminated
300-m dish. The main reflector consists of $\sim$4400 triangular
elements which allow surface formation from a sphere to a paraboloid
in real time via active control. The deep depression and feed cabin
suspension system allow a $40^\circ$ zenith angle, which may be
extended later by applying feeding techniques like Phased Array
Feeds (PAFs) in an upgrade stage \citep{nlj+11}. An order of
magnitude improvement in sensitivity can be expected with FAST
compared to a 100-m dish.

The Square Kilometre Array (SKA) with its collecting area of about
$10^6\,{\rm m}^2$ currently represents the ultimate design of future
radio telescopes \citep{sac+07}. It will be built in phases, but
already SKA Phase 1 will be amongst the largest radio telescopes on
Earth, and by far the largest on the southern hemisphere. The key
science of this international project is to address a wide range of
questions in astrophysics, fundamental physics, cosmology and
particle astrophysics, including gravitational wave astronomy in the
nano-Hz band and extreme GR tests including pulsar-black hole
binaries.

In the current SKA baseline design, the telescope consists of three
parts. For pulsar searches discussed here, the important components
are a sparse aperture array (SKA-low) of simple dipole antennas to
cover the low frequency range of 50--350\,MHz, and a dish array
(SKA-mid) of $\sim 15$\,m diameter elements to cover the high
frequency range from 350\,MHz to 14\,GHz. Each of them will
concentrate most of the collecting area in a central circular region
with diameter of about 5\,km. It is mostly this area that can be
phased-up for pulsar searches with synthesized beams within the
primary field-of-view.  While SKA1-low is supposed to provide a
sensitivity that is somewhat less than that of FAST, SKA1-mid should
provide similar sensitivity. The full SKA-mid in Phase II will be
equivalent to a telescope with an effective collecting area of about
one square kilometre, which will enable an $\sim50$ times
improvement in sensitivity at 1.4\,GHz compared with a 100\,m dish
\citep{sac+07}.

%------------------------------------------------------------------------------%

\subsection{Expected Improvements in pulsar timing precision}
\label{ssec:timing}

The current timing precision for slow pulsars is limited by
irregularities of the pulsar's spin \citep[e.g.][]{lhk+10}, which is
difficult to be improved by an increase in telescope sensitivity.
However, the rms timing residuals for most MSPs are still dominated
by system white noise \citep[e.g.][]{vbc+09}, which can be greatly
decreased by the increased instantaneous gain of the future
telescopes. Here, following the method in \citet{lvk+11}, we
estimated the expected measurement precision on pulse
time-of-arrivals (TOAs) at 1.4\,GHz with the application of the SKA
and the FAST, compared with that achievable with a 100-m dish. The
results together with the presumed instrumental sensitivities are
summarised in Table~\ref{tab:TOA_precision}. It can be seen that in
an optimal case the full SKA-mid will provide nearly two orders of
magnitude improvement to the timing precision. Note that we do not
consider intrinsic noise due to profile phase jitter, since firstly
the phenomenon is source dependent \citep[e.g.][]{jak+98,jg04},
secondly it can be decreased by extending integration time
\citep{cs10}, and thirdly it may be corrected with potential methods
being developed \citep[e.g.][]{ovh+11}.

\begin{table}
\centering \caption{Approximate values of the expected sensitivity
(effective collecting area, $A_{\rm eff}$, divided by system
temperature, $T_{\rm sys}$) at 1.4\,GHz and TOA precision with
10-min integration time $\sigma_{\rm 10min}$ for different
telescopes used in this paper \citep{nlj+11,dtm+13}. Here we assume
a pulsar of 1\,mJy flux density, 5\,ms period, and 100\,$\mu$s pulse
width. The observing bandwidth is assumed to be 500\,MHz.}
\begin{tabular}[c]{lcc} \\
\hline
&$A_{\rm eff}/T_{\rm sys}$ ($\rm m^2/K$) &$\sigma_{\rm 10min}$ ($\mu s$)\\
\hline
100-m dish  & 200 & 1.0  \\
FAST        & 2000 & 0.1 \\
SKA1-mid    & 1630 & 0.12 \\
Full SKA-mid    & 10000 & 0.02 \\
\hline
\end{tabular}
\label{tab:TOA_precision}
\end{table}

%%%%%%%%%%%%%%%%%%%%%%%%%%%%%%%%%%%%%%%%%%%%%%%%%%%%%%%%%%%%%%%%%%%%%%%%%%%%%%%%

\section{Testing the properties of a black-hole spacetime}
\label{sec:GR_test}

As an exact solution to Einstein's field equations of GR, the Kerr
metric describes the outer spacetime of an astrophysical (uncharged)
black hole \citep[e.g.][]{tpm86}. At the centre of the black hole
lies a gravitational singularity, a region where the curvature of
spacetime diverges. Penrose's ``Cosmic Censorship Conjecture''
states that within GR such singularities are always hidden within
the event of horizon \citep{pen79}, giving an upper limit for the
black hole spin $S_\bullet$, which is %%
\begin{equation}
  \chi \; \equiv \; \frac{c}{G}\,\frac{S_\bullet}{M^2_\bullet} \; \le 1 \;,
\end{equation}
where $c$ is the speed of light and $M_\bullet$ the mass of the
black hole. Therefore, measurements of the black hole mass and spin
can be used to test this inequality. A measured $\chi$ exceeding
unity in a tight pulsar binary system would pose an interesting
challenge for our understanding of the nature of the compact pulsar
companion, or even call our concept of gravity and spacetime into
question. Within the Kerr solution, for $\chi > 1$ the event horizon
vanishes, indicating the possibility of a spacetime singularity
being exposed to the outside universe, which is a violation of the
cosmic censorship conjecture \citep{pen79}. On the other hand, $\chi
> 1$ could signal an extremely unusual object within GR
\citep[e.g.][]{rya97}, or even the breakdown of GR itself.

Astrophysical black holes are believed to be results of a
gravitational collapse, during which all properties of the
progenitor, apart from the mass and spin, are radiated away by
gravitational radiation while the gravitational field asymptotically
approaches its stationary configuration \citep{pri72a,pri72b}.
Consequently, all higher multipole moments of the gravitational
field of an astrophysical black hole can be expressed as a function
of $M_\bullet$ and $S_\bullet$, which is the consequence of the
``no-hair theorem''\citep{han74}. In particular, the quadrupole
moment, $Q_\bullet$, fulfills the relation \citep{t80} %%
\begin{equation}\label{eq:q}
  q \; \equiv \; \frac{c^4}{G^2}\,\frac{Q_\bullet}{M^3_\bullet}
    \; = \; -\chi^2 \;,
\end{equation}
where $q$ is the dimensionless quadruple moment. A measurement of
$q$ in a clean system would therefore provide a (model-independent)
test of the no-hair theorem for a Kerr black hole. In the future,
this could be achieved, for instance, through the observations of
gravitational waves from the inspiral of a compact object into a
massive black hole \citep{rya95}, or the timing observations of a
pulsar in a tight orbit around a black hole
\citep{wk99,kbc+04,lwk+12}.

Concerning timing observations of a PSR-SBH system, the mass
determination can be achieved by measuring a variety of relativistic
effects that can be parametrized by a set of post-Keplerian (PK)
parameters, just like the previous experiments with binary pulsars
\citep[see][ for a detailed summary]{lk05}. For a Kerr black hole,
the spin-induced frame dragging will cause the pulsar orbit to
precess about the direction of the total angular momentum, which can
be properly modelled and used to derive the spin and system
geometry. The quadrupolar potential due to the ``oblateness'' of the
black hole will induce periodic perturbations to the pulsar orbit,
which in principle may lead to a measurement of the black hole
quadrupole moment.

This section will begin with an extensive discussion on the mass
determination for pulsar and black hole, and then investigate in
detail the effects of the black-hole spin and quadrupole moment on
the orbital dynamics of the pulsar and their measurability through
pulsar timing.

%------------------------------------------------------------------------------%

\subsection{Mass measurement} \label{ssec:mass}

The mass measurement for the black hole is essential in identifying
the pulsar companion as a black hole. A well determined mass that
clearly exceeds the maximum mass of a neutron star ($\lesssim
3\,M_\odot$) for any reasonable equation-of-state (EoS), in
combination with optical observations\footnote{If the pulsar is in a
tight orbit around its massive companion, then already the absence
of eclipses and tidal effects would argue for a sufficiently compact
companion, and consequently for a black hole.} that can exclude a
main-sequence star, would make a very strong case for the black-hole
nature of the pulsar companion. In principle, there could be more
exotic alternatives to black holes, but many of them are expected to
differ in their spin and quadrupole properties, a test where again a
precise mass is the key input.

When describing timing observation of relativistic pulsar binary
systems, one can use a set of theory-independent PK parameters
which, for a given theory of gravity, are theory-specific functions
of the two a priori unknown masses \citep[e.g.][]{dd86,dt92}.
Consequently, if measurements of any two PK parameters are achieved,
the masses can then be derived assuming GR is correct. The measurement
of a third PK parameter would then verify the applicability of GR for
the mass determination. In the
following, we briefly introduce several such parameters that are
commonly used in pulsar timing for mass determinations, and discuss
their measurability in PSR-SBH systems, based on mock data
simulations. We keep only the leading terms which are sufficient to
estimate the measurement precisions, as argued in \citet{lwk+12}.

\subsubsection{Post-Keplerian Parameters and Mass Determination} \label{sssec:PK}

In eccentric pulsar binaries, the precession of periastron is
usually the first PK parameter that can be measured with high
precision. Typically, the major contribution is from the mass
monopoles which following \citet{rob38}, can be written as %%
\begin{equation}\label{eq:PKomdot}
  \dot{\omega}_{\rm m} =
    \frac{3}{1-e^2}\left(\frac{P_{\rm b}}{2\pi}\right)^{-5/3}
    T_{\odot}^{2/3} \, (m_{\rm p}+m_\bullet)^{2/3} \;,
\end{equation}
and for $m_\bullet \gtrsim 10$ be approximated by %%
\begin{equation}
  \dot{\omega}_{\rm m} \approx
    (\rm 0.92\,{\rm deg\,yr^{-1}})\,\frac{1}{1-e^2}
    \left(\frac{P_{\rm b}}{\rm 1\,day}\right)^{-5/3}
    \left(\frac{m_\bullet}{10}\right)^{2/3} \;,
\end{equation}
where $e$ is the orbital eccentricity, $P_{\rm b}$ is the orbital
period, $T_\odot \equiv GM_\odot/c^3 \simeq 4.9255\,\mu$s is the
solar mass in second, $m_{\rm p}$ and $m_\bullet$ are masses of the
pulsar and the companion black hole in solar units, respectively.
For systems of $m_\bullet > 10$ and $P_{\rm b} < 1$\,day, the
precession rate is larger than $\rm 1\,deg\,yr^{-1}$, which after a
few years of timing observations would result in a precise
measurement of $\dot{\omega}$. However, if the black hole in a
PSR-SBH system is fast rotating, a significant fraction of
periastron advance can also be induced by frame dragging
\citep{bo75b}. In this case, the observed $\dot{\omega}$ cannot be
used directly for a precise mass determination, but is useful in determining
the black hole spin and system geometry, which will be shown in
Section~\ref{ssec:spin}.

The {\em Einstein delay} is a combination of the second order
Doppler effect and gravitational redshift. Its amplitude is also a
PK parameter and within GR determined as \citep{bt76} %%
\begin{eqnarray}
  \gamma &=& e \left(\frac{P_{\rm b}}{2\pi}\right)^{1/3} T_{\odot}^{2/3} \,
    \frac{(m_{\rm p} + 2m_\bullet)m_\bullet}{(m_{\rm p}+m_\bullet)^{4/3}}
  \nonumber\\
  &\approx& (64\,{\rm ms})\,e
    \left(\frac{P_{\rm b}}{\rm 1\,day}\right)^{1/3}
    \left(\frac{m_\bullet}{\rm 10}\right)^{2/3},
    \; \mbox{for $m_\bullet\gg m_{\rm p}$} \;.
  \label{eq:PKgamma}
\end{eqnarray}
Clearly, the effect is important only when the system is
sufficiently eccentric (e.g. $e\gtrsim 0.1$). For eccentric orbits with
$P_{\rm b} \sim 0.1$ to 1\,days and $m_\bullet \gtrsim 10$, $\gamma$
would be of order $10 \sim 100$\,ms, well above the expected timing
precision. However, at the beginning of observation the Einstein
delay is always degenerate with the R\"omer delay\footnote{ The
R\"omer delay is defined as $\Delta_{\rm R} = \hat{\bf K}_0 \cdot
{\bf r}$, where $\hat{\bf K}_0$ is the unit vector along the of the
line-of-sight to the PSR-SBH system, and {\bf r} is the position
vector of the pulsar with respect to the barycenter of the binary
system. It describes the contribution of the pulsar's orbital motion
to the signal travel time.} and is separable only when the
relativistic advance of pericentre (see Eq.~\ref{eq:PKomdot})
changed the orbital orientation by a sufficient amount.

The {\em Shapiro delay} accounts for the additional light travel
time due to the curvature of spacetime induced by the companion
mass. It contains two separately measurable PK parameters, which in
GR read %%
\begin{equation} \label{eq:PKshap}
  r_{\rm Sh} = T_{\odot} m_\bullet \;, \quad s_{\rm Sh} = \sin i \;.
\end{equation}
The angle $i$ is defined as the angle between the line-of-sight and the
orbital angular momentum of the binary system. Following \cite{bt76}
and \cite{dd86} one finds: %%
\begin{eqnarray}
  \Delta_{\rm Sh} &=& 2 \, r_{\rm Sh} \,
  \ln\left[\frac{1 + e\cos\phi}{1 - s_{\rm Sh}\sin(\omega + \phi)}\right]
  \nonumber\\
  &\simeq& (98.5\,\mu{\rm s})
  \left(\frac{m_\bullet}{\rm 10}\right)
  \ln\left[\frac{1+e\cos\phi}{1 - s_{\rm Sh} \sin(\omega +\phi)}\right] \;,
  \label{eq:DeltaS}
\end{eqnarray}
where $\phi$ is the orbital true anomaly. For systems of either
significant eccentricity or inclination angle, the signal would be
of order $10\sim100$\,$\mu$s when $m_\bullet\gtrsim10$. This is
still well above the expected timing precision and should allow
precise mass measurements. In GR, $\sin i$ is related to $m_{\rm p}$
and $m_\bullet$ by the mass function: %%
\begin{equation} \label{eq:PKs}
  s_{\rm Sh} \equiv \sin i = x \left(\frac{P_{\rm b}}{2\pi}\right)^{-2/3}\,
    T_{\odot}^{-1/3} \frac{(m_{\rm p}+m_\bullet)^{2/3}}{m_\bullet} \;,
\end{equation}
where $x$ is the projected semi-major axis of the pulsar orbit
measured in light-seconds, i.e.\ $x \equiv a_{\rm p} \sin i/c$.

The change in orbital period is an additional PK parameter directly
measurable from pulsar timing. In GR, the quadrupole radiation
predicts \citep[e.g.][]{pet64} %%
\begin{equation}
  \dot{P}_{\rm b} = -\frac{192\pi}{5}
    \left(\frac{P_{\rm b}}{2\pi}\right)^{-5/3}
    T_{\odot}^{5/3} \,
    \frac{m_{\rm p}m_\bullet}{(m_{\rm p}+m_\bullet)^{1/3}} \, f(e) \;,
\end{equation}
where %%
\begin{equation}
  f(e) = \frac{1 + (73/24)e^{2} + (37/96)e^{4}}{(1 - e^{2})^{7/2}} \;.
\end{equation}
For $m_\bullet \gtrsim 10$ and $m_{\rm p} \approx 1.4$ one has %%
\begin{equation}
  \dot{P}_{\rm b} \approx (-4.5\,\mu{\rm s}\,{\rm yr}^{-1})
    \left(\frac{P_{\rm b}}{\rm 1\,day}\right)^{-5/3}
    \left(\frac{m_\bullet}{\rm 10}\right)^{5/3} f(e) \;.
\end{equation}
Consequently, in PSR-SBH systems with short orbital periods ($\sim
1$\,day), $\dot{P}_{\rm b}$ should be measurable with high precision
after only a few years of timing observations.

\subsubsection{Mock data analysis}
\label{sssec:masssimu}

The measurability of the PK parameters in PSR-SBH systems have been
investigated based on simulated timing data. The simulations
performed in this paper mainly contain two steps. Firstly, based on
presumed observing scheme, the TOAs are generated regularly
regarding to pulsar's initial time and then combined with the three
time delays (R\"omer, Einstein and Shapiro) to account for the
changes in the signal arrival time at the barycentre due to the
pulsar's orbital motion. Next, the simulated TOAs are passed to the
TEMPO software package, which, based on a timing model, performs a
least-square fit to yield a phase-connected solution of the TOAs,
and determines the timing model parameters. The measurement
uncertainties of these parameters are produced from the covariance
matrix. In this subsection we used the DD timing model \citep{dd86}
to obtain the following results.

\begin{figure}
\centering
\includegraphics[scale=0.6]{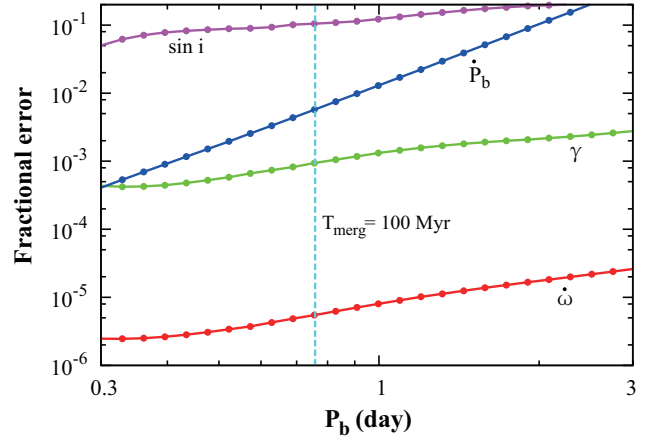}
\caption{Simulated fractional measurement errors of PK parameters as
a function of $P_{\rm b}$ for PSR-SBH systems of a slow pulsar. The
assumed system parameters are: $e=0.8$, $m_{\rm p}=1.4$, $m_\bullet
=10$, $i=60^\circ$. Here we assume ten TOAs per week of 100\,$\mu$s
precision for timing observation with 5-yr baseline. The simulation
is carried out for non-rotating black holes. \label{fig:mass-norm}}
\end{figure}

In Fig.~\ref{fig:mass-norm} we present expected measurement
precisions of the PK parameters from PSR-SBH systems where the
pulsar is non-recycled. Here as mentioned in
Section~\ref{sec:intro}, the orbit is likely to be highly eccentric.
Note that the timing precision of slow pulsars are normally
dominated by timing noise and thus not expected to be significantly
improved by system sensitivity \citep[e.g.][]{lhk+10}. Accordingly,
we assumed 5-yr observations leading to ten TOAs per week with a
precision of 100\,$\mu$s. To indicate systems with a realistic
lifetime, we label the orbital period corresponding to a
gravitational merging timescale of 100\,Myr, the value of the most
relativistic (currently known) binary pulsar system PSR~J0737$-$3039.
The merging timescale is given by \citep[e.g.][]{pet64} %%
\begin{equation}\label{eq:T_merg}
  T_{\rm merg} =
    \frac{5}{256}\,\left(\frac{P_{\rm b}}{2\pi}\right)^{8/3}
    T^{-5/3}_{\odot}
    \frac{(m_{\rm p} + m_\bullet)^{1/3}}{m_{\rm p}m_\bullet}\,g(e) \;,
\end{equation}
where %%
\begin{eqnarray}
  g(e) &=& 1 - 3.6481\,e^2 + 5.1237\,e^4 - 3.5427\,e^6
  \nonumber\\&&\qquad\qquad\qquad\qquad
                  + 1.3124\,e^8 - 0.2453\,e^{10}
\end{eqnarray}
represents an approximation to the corresponding integral in
Eq.~(5.14) of \cite{pet64}, with a $<1$\% fractional error for $e
\le 0.9$. For $m_\bullet \gtrsim 10$ and $m_p \approx 1.4$ one finds
\begin{equation}
  T_{\rm merg} \approx (725\,{\rm Myr})
    \left(\frac{P_{\rm b}}{\rm 1\,day}\right)^{8/3}
    \left(\frac{m_\bullet}{\rm 10}\right)^{-2/3} g(e) \;.
\end{equation}
It can be seen that after five years of observation, the masses can be
obtained with high precision if the orbital period is of a few days or less. The
measurable PK parameters are most likely to be $\dot{\omega}$,
$\gamma$, and $\dot{P}_{\rm b}$.

\begin{figure}
\centering
\includegraphics[scale=0.6,angle=-90]{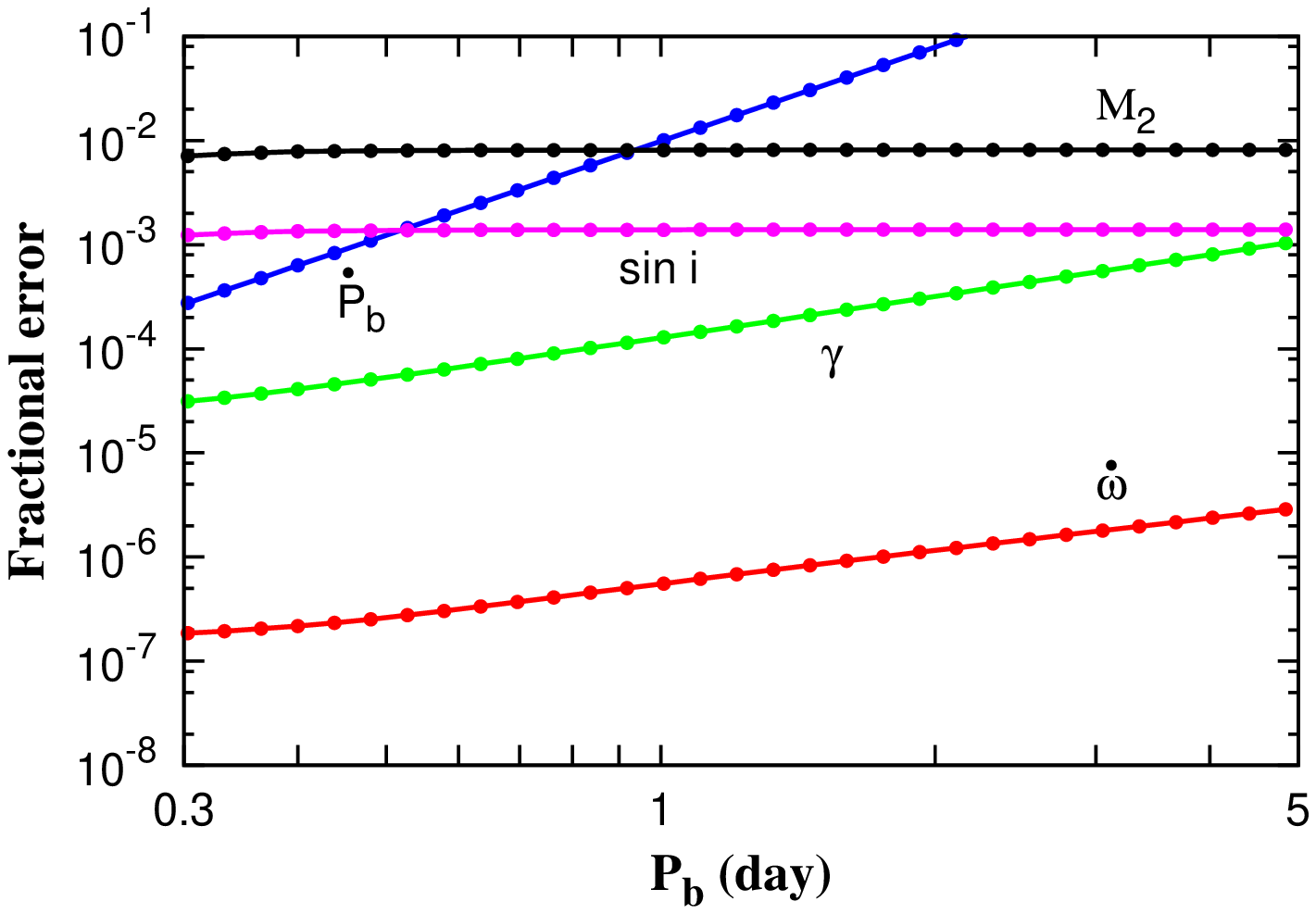}
\includegraphics[scale=0.6,angle=-90]{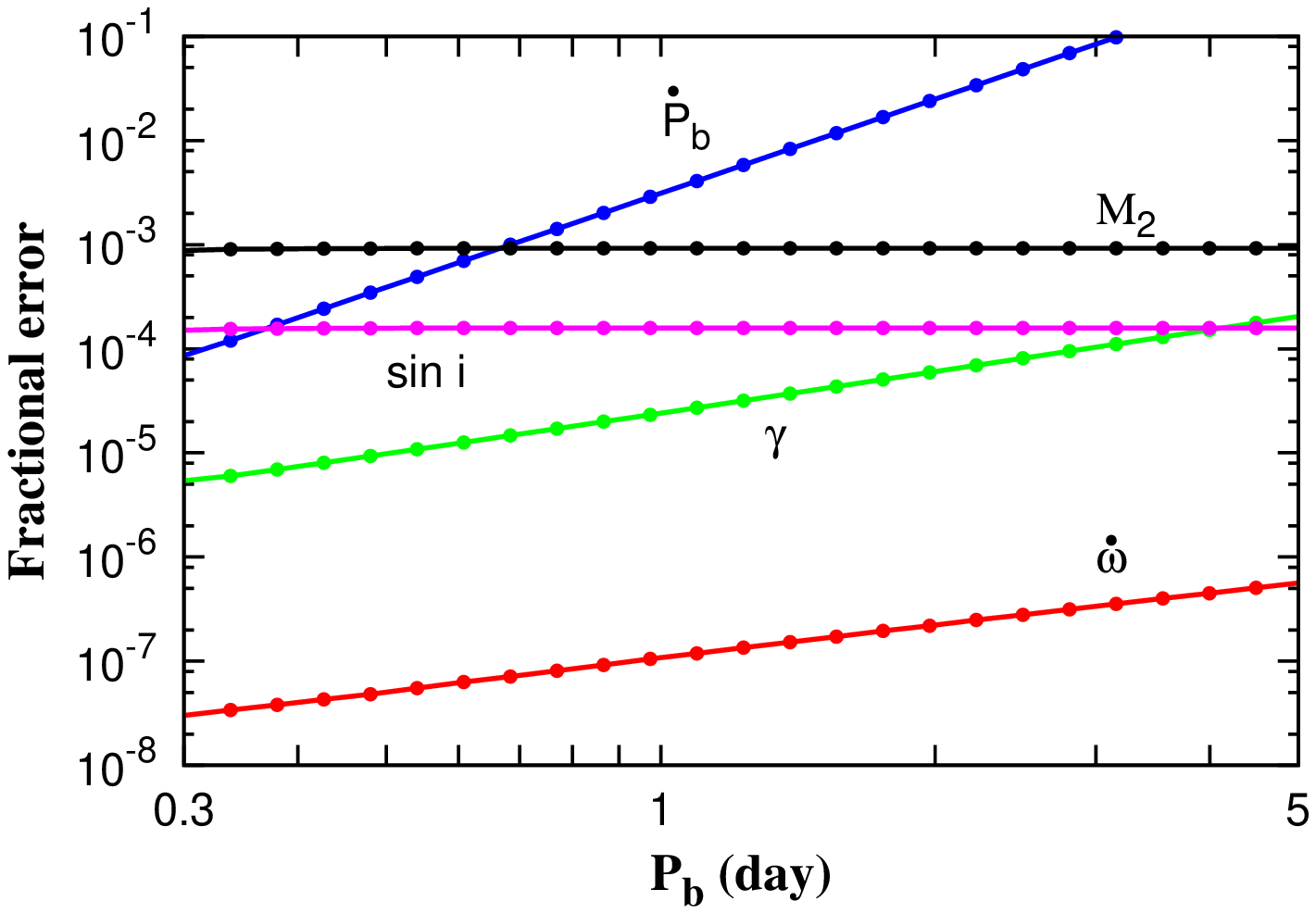}
\includegraphics[scale=0.6,angle=-90]{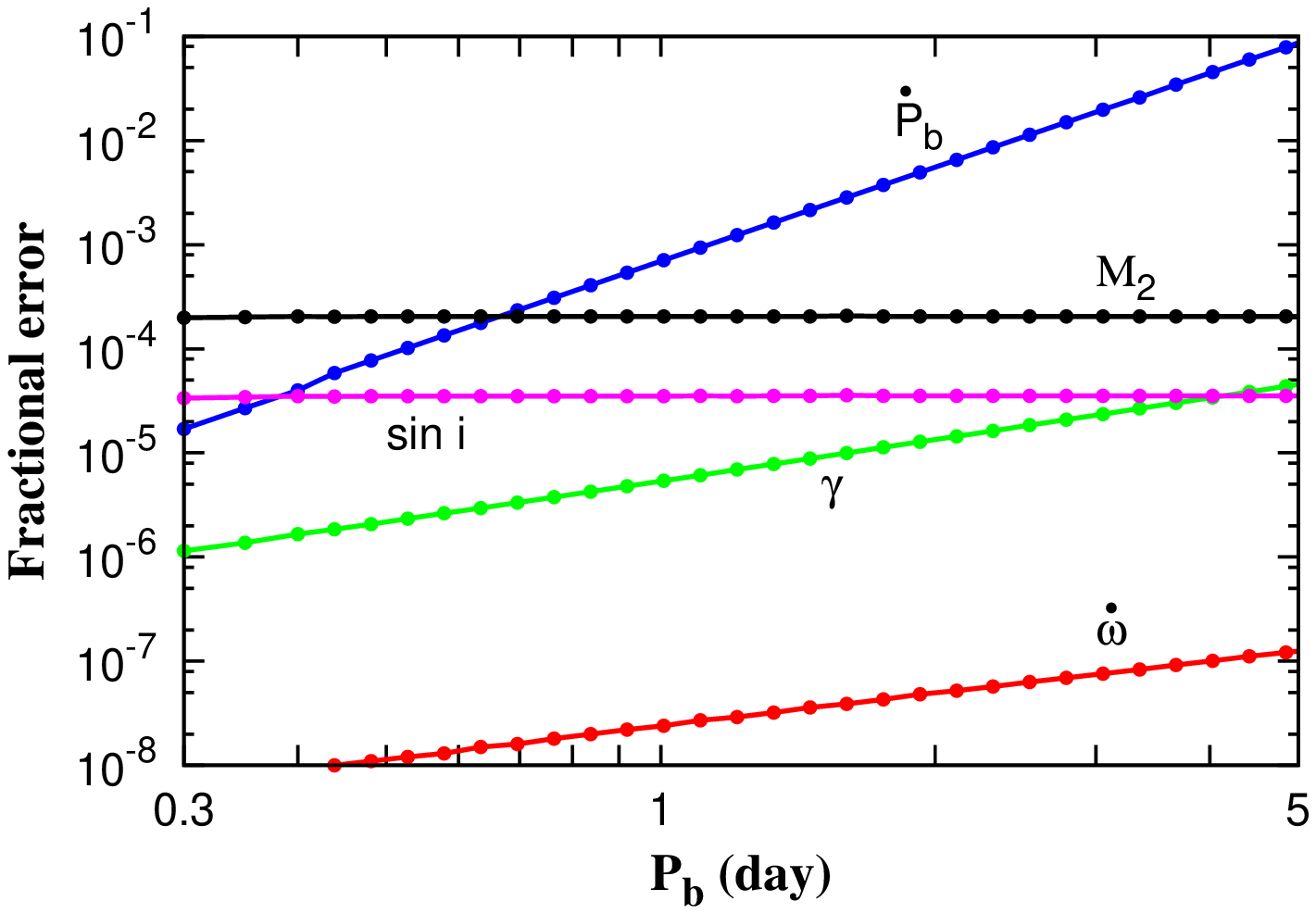}
\caption{Simulated fractional measurement errors of PK parameters as
a function of $P_{\rm b}$ for PSR-SBH systems with a MSP. Here we use
$e=0.1$, $m_{\rm p}=1.4$, $m_\bullet =10$, $i=60^\circ$. Weekly four
hour timing observations are assumed, for five years long (top plot)
with a 100-m dish and for three years long (middle plot for the
FAST, bottom plot for the SKA) for the future telescopes.
\label{fig:mass-e0.1}}
\end{figure}

In Fig.~\ref{fig:mass-e0.1} we consider PK parameter measurements
for PSR-SBH systems with a MSP. Here we applied 4-hr observations per
week with the sensitivity of a 100-m dish (top), the FAST (middle)
and SKA (bottom), which can be converted into ten TOAs with
precision of 1\,$\mu$s, 100\,ns and 20\,ns, respectively. The timing
baselines were assumed to be 5\,yr for the 100-m dish and 3\,yr for
the future telescopes. It is shown that with the sensitivity of a
current 100-m dish after five years of observation the masses are
likely to be measured with precision better than 1\%. The
measurement precision is expected to be improved by a factor of 10
to 100 with the next generation of radio telescopes.

Note that in Fig.~\ref{fig:mass-e0.1} we used a low eccentricity of
$e=0.1$. If a more eccentric system is found, given the same orbital
period, the relativistic effects would be stronger thus measured
with better precision. In addition, as indicated from
Eq.~(\ref{eq:DeltaS}), the measurability of the Shapiro delay
parameters are not significantly dependent on the orbital period.
Thus, their measurements would enable mass determination for systems
of wide orbits (e.g., $P_{\rm b} \sim 100$\,days). In either case,
the periastron advance is well measured but as mentioned in
Section~\ref{sssec:PK}, can be directly used for mass measurements
only when the black hole is not significantly spinning or the orbit is sufficiently wide.

%------------------------------------------------------------------------------%

\subsection{Frame dragging, spin measurement and cosmic censorship conjecture} \label{ssec:spin}

In a binary system, additional to the orbital precession induced by
mass monopoles, the spin of the bodies will drag the spacetime in the
vicinity and cause an extra precession to the orbit
\citep{lt18,wex95}. It was demonstrated by \citet{wk99} that in a
PSR-SBH system, the orbital precession can be measured through pulsar timing
observations, and
used to determine the black hole spin. In this subsection we will
present a brief description of this effect, and based on mock data
simulations, investigate the measurability of the black hole spin
and test of the cosmic censorship conjecture with the next
generation of radio telescopes.

\subsubsection{Orbital precession and its consequence on timing}
\label{sssec:fd}

\begin{figure}
\centering
\includegraphics[scale=0.5]{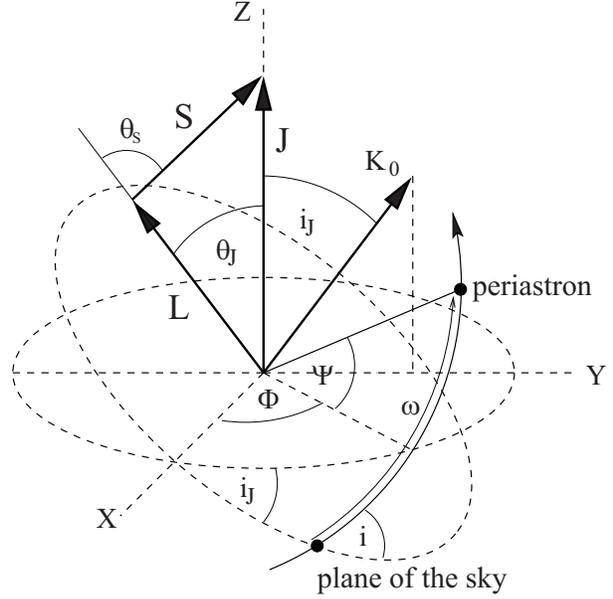}
\caption{Geometry of a PSR-SBH system. The reference frame is based
on the invariable plane perpendicular to the system total angular
momentum $\vec{J}$. The line-of-sight vector $\vec{K}_0$ is fixed to
the Y-Z plane, while the orbital momentum $\vec{L}$ is supposed to
precess around $\vec{J}$. The definition of angle $\theta_S$, $i_J$,
$\Phi$ and $\Psi$ will present a full description of the orbital
geometry. The corresponding defined ranges are:
$\theta_S,~\theta_J,~i_J,~i\in[0,~\pi)$ and
$\Phi,~\Psi,~\omega\in[0,~2\pi)$. \label{fig:Jcoord}}
\end{figure}

The precession due to frame-dragging can be best described in a
coordinate based on the invariant plane perpendicular to the total
angular momentum $\vec{J}$, as shown in Fig.~\ref{fig:Jcoord}. In
general, $\vec{J}$ can be considered as a conserved quantity and
both the orbital angular momentum, $\vec{L}$, and the black hole
spin, $\vec{S}$, are supposed to precess around $\vec{J}$. Their
absolute values are also conserved, if averaged over a whole
orbital period. Following \citet{bo75b} one finds the orbit-averaged
frame-dragging precession rate in the form of
\begin{eqnarray}
  \dot{\Phi}_S &=&
    +\Omega^{*}_S\,\frac{\sin\theta_S}{\sin\theta_J} \;,
    \label{eq:Phidot}\\
  \dot{\Psi}_S  &=&
    -\Omega^{*}_S\,(2\cos\theta_S + \sin\theta_S\cot\theta_J) \;,
    \label{eq:Psidot}
\end{eqnarray}
where %%
\begin{equation}
 \Omega^{*}_S  = \frac{\chi}{(1 - e^2)^{3/2}}
   \left(\frac{2\pi}{P_{\rm b}}\right)^2 T_{\odot} \,
   \frac{(3m_{\rm p} + 4m_\bullet)m_\bullet}{2(m_{\rm p} + m_\bullet)}\;.
\end{equation}
For $m_\bullet \gtrsim 10$ and $m_{\rm p} \approx 1.4$ one finds the
following approximate expression %%
\begin{equation}
  \Omega^{*}_S \approx(4.4 \times 10^{-3}\,{\rm deg\,yr^{-1}}) \,
    \frac{\chi}{(1 - e^2)^{3/2}} \,
    \left(\frac{P_{\rm b}}{\rm 1\,day}\right)^{-2}
    \left(\frac{m_\bullet}{\rm 10}\right) \;.
\end{equation}
The linear-in-time secular changes in $\Phi$ and $\Psi$ induce
non-linear-in-time evolution in two timing parameters, the longitude
of periastron $\omega$ and projected semi-major axis $x$. In real
timing observations, the variations can be approximated by Taylor
expansions as below: %%
\begin{eqnarray}
  x     &=& x_0 + \dot{x}_0(t - T_0) + \frac{1}{2}\ddot{x}_0(t - T_0)^2 + \dots\;,\\
 \omega &=& \omega_0 + \dot{\omega}_0 (t - T_0) +
            \frac{1}{2}\ddot{\omega}_0(t - T_0)^2 + \dots\;,
\end{eqnarray}
where $T_0$ is the time of periastron passage and $x_0$, $\omega_0$
denote the initial values at $T_0$. As mentioned in
Section~\ref{sssec:PK}, the observed $\dot{\omega}_0$ would have a
significant contribution caused by the mass monopoles.
The contribution by frame-dragging
can be determined by calculating the monopole component from mass
measurements by the other PK parameters and subtract it from the
observed $\dot{\omega}_0$.

The contributions to the first and second derivatives, arising from
the frame dragging of the black hole companion, have been worked out
in detail by \citet{wk99}\footnote{In Eq.~(64) of \cite{wk99} there
is a sign error, $\Omega_S^\ast$ has to be replaced by
$-\Omega_S^\ast$.}: %%
\begin{eqnarray}
  \dot{x}_S  &\simeq& -x_0 \, \chi \tilde{\Omega} \,
    \cot i \sin\theta_S \sin\Phi_0 \;,
    \label{eq:SMBxdot} \\
  \ddot{x}_S &\simeq& -x_0 \, \chi \tilde{\Omega}^2 \, \tilde{\Sigma}^{-1}
    \cot i \sin\theta_S \cos\Phi_0 \;,
    \label{eq:SMBx2dot} \\
  \dot{\omega}_S &\simeq& -\chi \tilde{\Omega} \,
    (2\cos\theta_S + \cot i\sin\theta_S\cos\Phi_0) \;,
    \label{eq:SMBomdot} \\
  \ddot{\omega}_S &\simeq& +\chi \tilde{\Omega}^2 \, \tilde{\Sigma}^{-1}
    \cot i \sin\theta_S \sin\Phi_0 \;,
    \label{eq:SMBom2dot}
\end{eqnarray}
where %%
\begin{equation}\label{eq:tildeOMs}
  \tilde{\Omega} \equiv \Omega^{*}_S/\chi
\end{equation}
corresponds to $\Omega^{*}_S$ of an extreme Kerr black hole ($\chi =
1$), and %%
\begin{equation}\label{eq:SLR}
   \tilde{\Sigma} \equiv \frac{S_\bullet/\chi}{L} =
    \left(\frac{P_{\rm b}}{2\pi}\right)^{-1/3}  T^{1/3}_{\odot} \,
    \frac{(m_{\rm p} + m_\bullet)^{1/3}m_\bullet}{m_{\rm p}
    (1 - e^2)^{1/2}} \;.
\end{equation}
For $m_\bullet \gtrsim 10$ and $m_{\rm p} \approx 1.4$ one finds %%
\begin{equation}
  \tilde{\Sigma} \approx 0.017
      \left(\frac{P_{\rm b}}{\rm 1\,day}\right)^{-1/3}
      \left(\frac{m_\bullet}{\rm 10}\right)^{4/3} \;.
\end{equation}
Therefore, once these four derivatives are measured, they can then
be used to derive $\chi$, $\theta_S$, $\Phi_0$ via
Eq.~(\ref{eq:SMBxdot})--(\ref{eq:SMBom2dot})\footnote{Note that
$\dot{x}_S$ and $\ddot{\omega}_S$ have the same dependency on spin
and system geometry. Here we use $\dot{x}_S$ since it is generally
measured with much higher precision, as will be shown in
Section~\ref{sssec:spin simu}.}, which gives the spin amplitude
$\chi$ and orientation $\theta_S$ as %%
\begin{equation} \label{eq:chi-pbhb}
  \chi \simeq \frac{1}{x_0\tilde{\Omega}}
    \left[ \left(\frac{ \Xi_S - x_0\dot{\omega}_S}{2}\right)^2 +
    \left( \dot{x}_S^2 + \Xi_S^2 \right) \tan^2i
    \right]^{1/2}
\end{equation}
and %%
\begin{equation}
  \cos\theta_S \simeq
    \frac{\Xi_S - x_0 \dot{\omega}_S}
    {
    \left[(\Xi_S - x_0\dot{\omega}_S)^2 +
          4(\dot{x}^2_S+\Xi^2_S)\tan^2i\right]^{1/2}}
\end{equation}
respectively, where %%
\begin{equation}
  \Xi_S\equiv\ddot{x}_S\tilde{\Sigma}/\tilde{\Omega} \;.
\end{equation}
$\theta_S$ is uniquely determined from $\cos\theta_S$ with the given
range of 0 to $\pi$. Once the spin is determined, the angles in
Fig.~\ref{fig:Jcoord} are all known at the same time. The
$i\leftrightarrow\pi-i$ ambiguity from $\sin i$ leads to two
different solutions in the orientation by $(\Phi_0,
i_J)\leftrightarrow(\pi+\Phi_0, \pi-i_J)$.

\subsubsection{Mock data analysis}
\label{sssec:spin simu}

Based on the simulation scheme described in
Section~\ref{sssec:masssimu}, we have estimated the measurability of
the black hole spin by timing a PSR-SBH system. Here when
calculating the time delays we also include the orbital precession
due to the black hole spin, by inputting the secular changes of
$\Phi$ and $\Psi$ described in
Eq.~(\ref{eq:Phidot})--(\ref{eq:Psidot}). Then we fit the calculated
TOAs with the MSS timing model of TEMPO \citep{wex98} to determine
the black hole spin and its orientation, as described in
\cite{lwk+12}.

\begin{figure}
\centering
\includegraphics[scale=0.6]{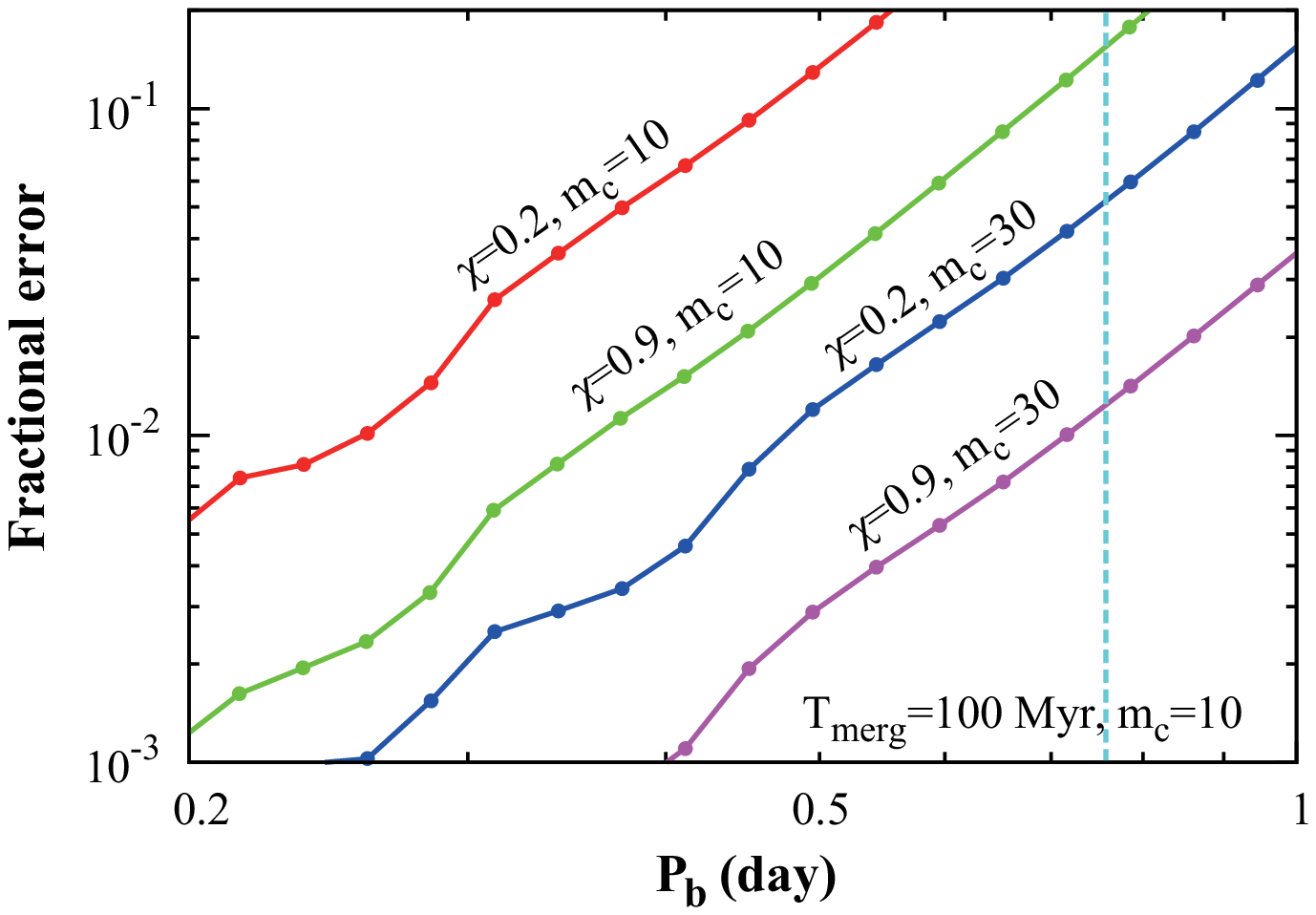}
\includegraphics[scale=0.6]{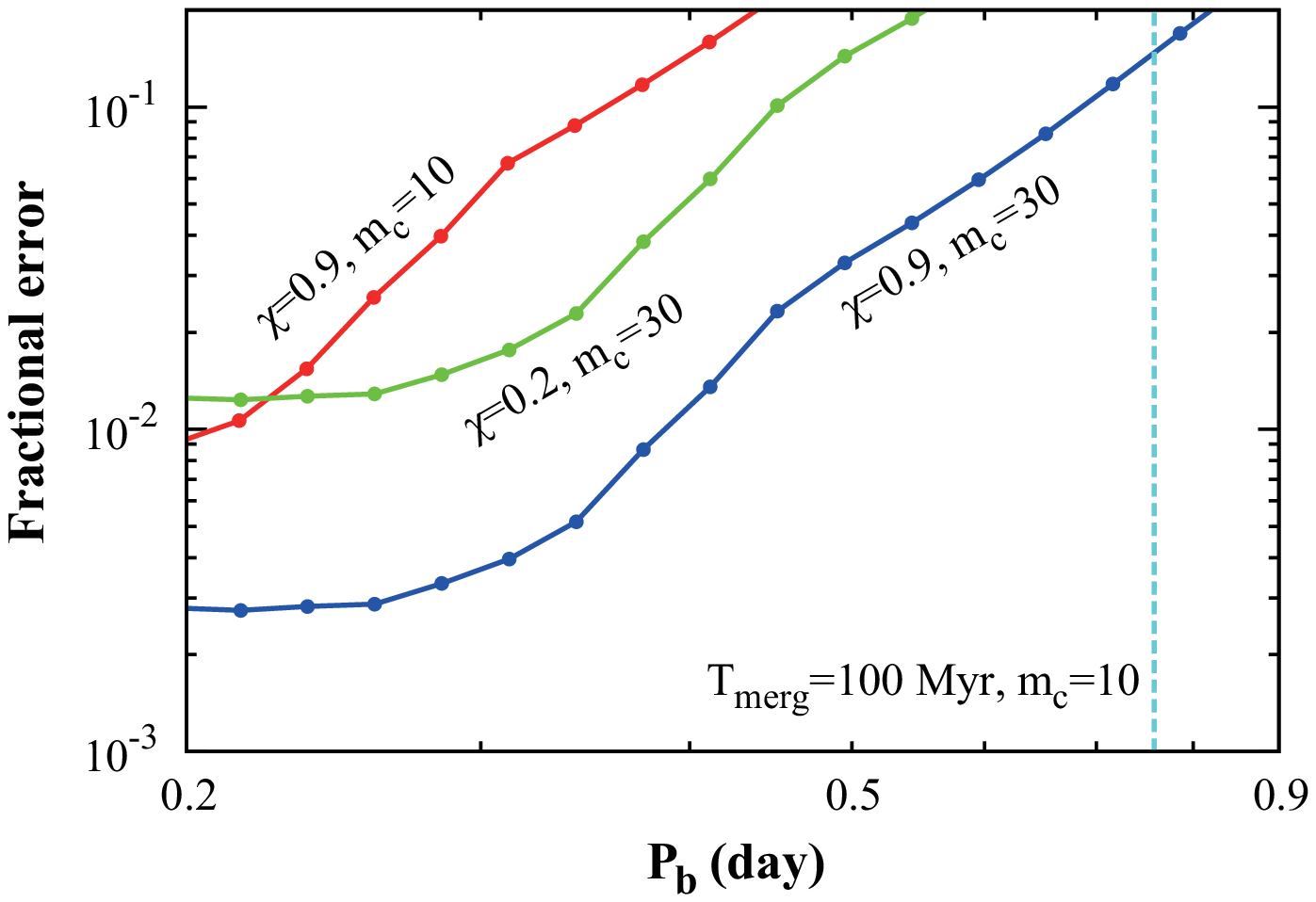}
\caption{Simulated fractional measurement errors of frame the dragging
effect as a function of orbital period for PSR-SBH systems with a
slow pulsar and a 10 and 30\,$M_\odot$ BH.
The system parameters are: $e = 0.8$, $m_{\rm p} =
1.4$, $\theta_S = \Phi_0 = 45^{\circ}$, $i = \Psi_0 = 60^{\circ}$.
Here we apply the same weekly observational scheme as in
Fig.~\ref{fig:mass-norm}, with a baseline of ten years. The top plot
shows the measurement precision of secular change in orbital
projected semi-major axis, the first sign for the existence of black
hole spin. The bottom plot presents the measurability of the spin
magnitude. \label{fig:spin-norm}}
\end{figure}

The expected black hole spin measurement precision in PSR-SBH
systems of a slow pulsar is investigated in
Fig.~\ref{fig:spin-norm}. Here we also show the measurability of the
secular change in orbital projected semi-major axis (top), which
usually provides the first sign of the black hole spin. We applied
the same weekly observing scheme as described in
Fig.~\ref{fig:mass-norm}, but extend the timing baseline to ten
years. The results suggest that by timing a slow pulsar for ten
years, the spin determination is achievable for a wide range of
$m_\bullet$ and $\chi$ only if the orbit is compact enough ($P_{\rm
b}\lesssim 0.5$\,days) and highly eccentric ($e \gtrsim 0.8$). In wide orbits
($P_{\rm b}\gtrsim 1$\,day),
the measurement can be achieved only if the black hole is comparably
more massive and fast rotating. Nevertheless, the sign of the frame
dragging effect ($\dot{x}$) can be noticed for a wide range of BH masses
and spins when $P_{\rm b} \lesssim 1$\,day.

\begin{figure}\color{white}[]
\centering \color{black}
\includegraphics[scale=0.6,angle=-90]{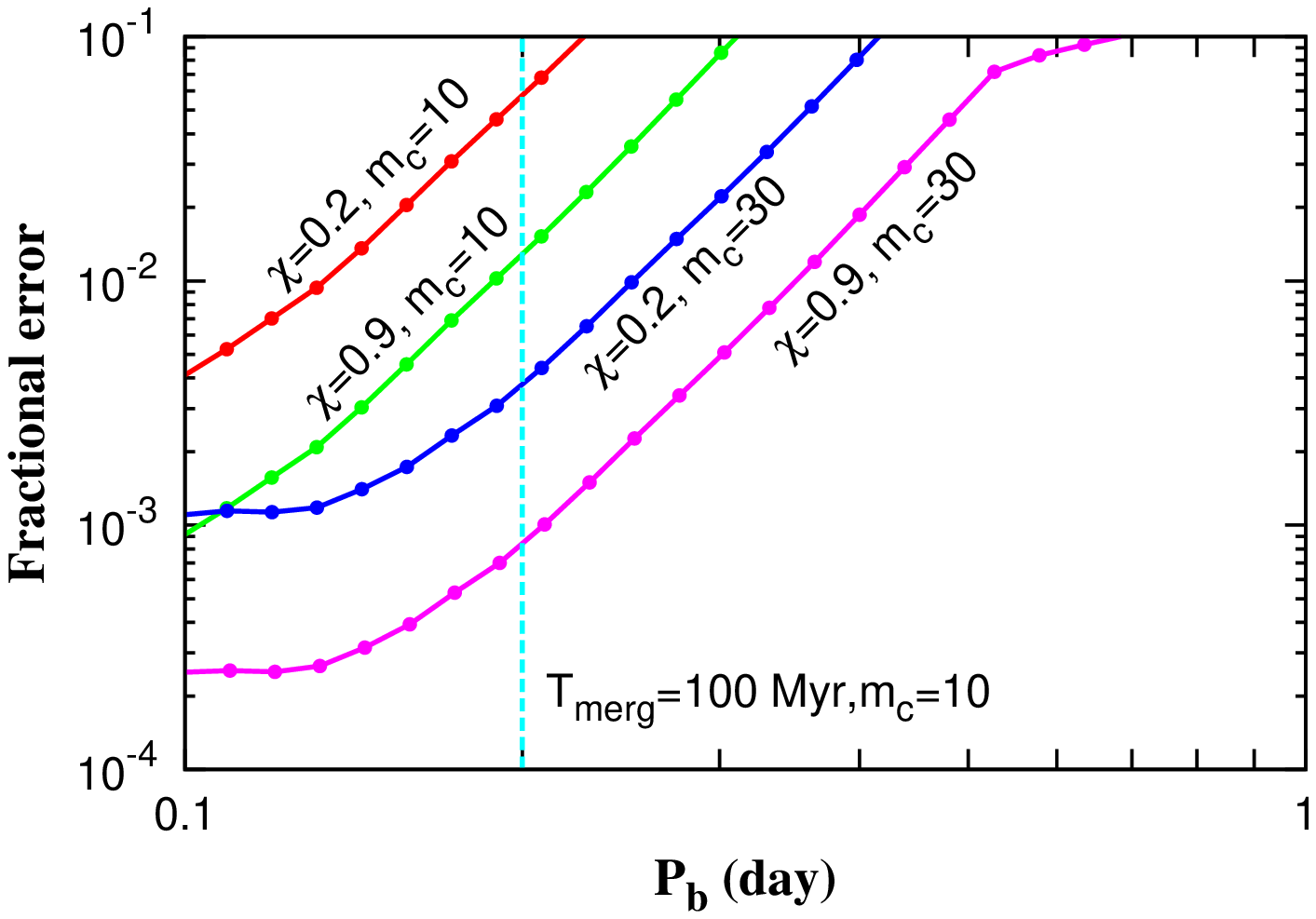}
\includegraphics[scale=0.6,angle=-90]{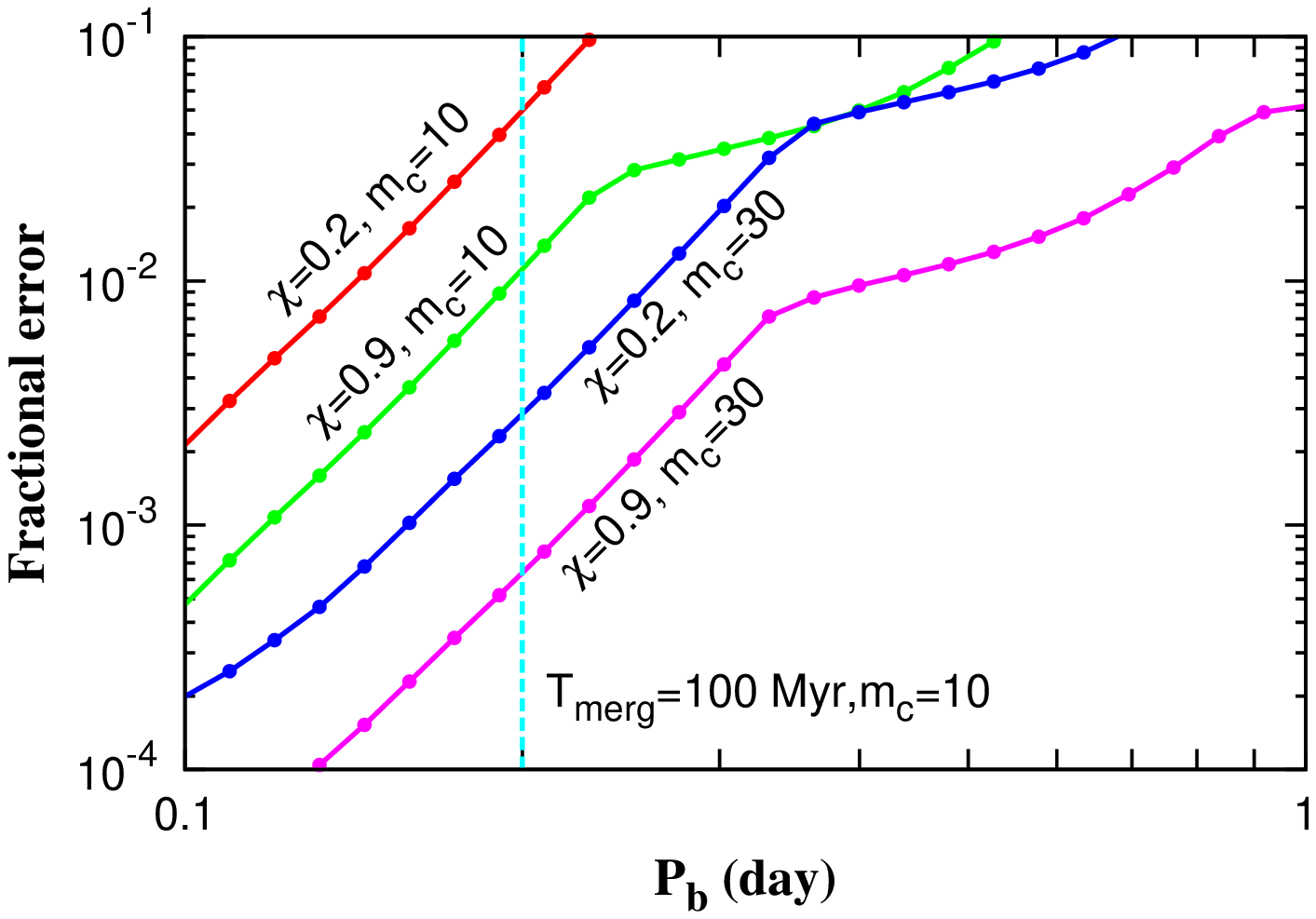}
\includegraphics[scale=0.6,angle=-90]{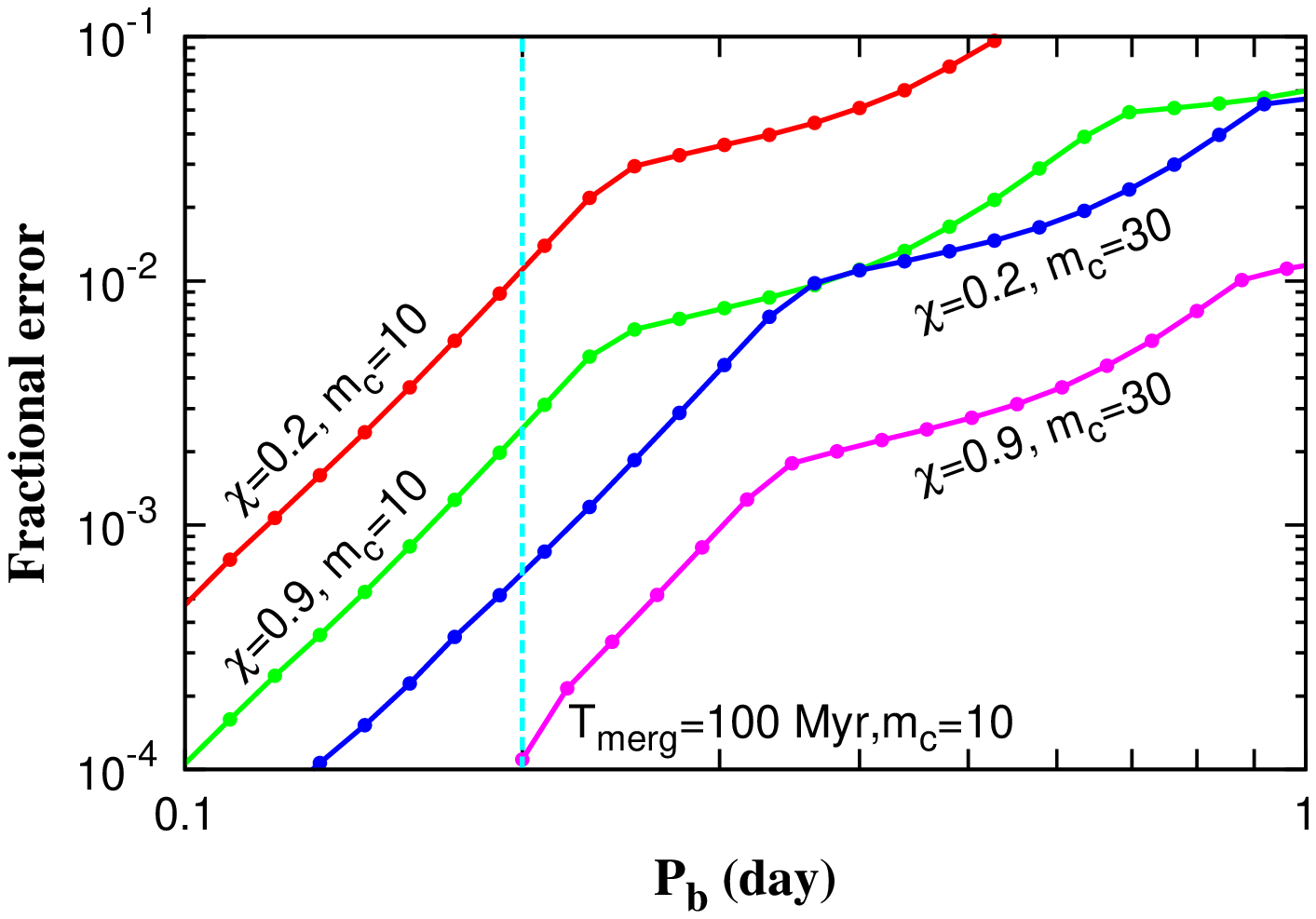}
\caption{Simulated fractional measurement errors of black hole spin
as a function of orbital period for PSR-SBH systems with a MSP.
Herethe system parameters are: $e=0.1$, $m_{\rm p} = 1.4$, $\theta_S =
\Phi_0 = 45^{\circ}$, $i = \Psi_0 = 60^{\circ}$.
For the BH mass we assume 10 and 30\,$M_\odot$.
The observations
were assumed to be of the same weekly scheme as in
Fig.~\ref{fig:mass-e0.1}, with baseline of 10\,yr for a 100-m dish
sensitivity in the top plot and 5\,yr for the future telescopes
sensitivity in the middle (FAST) and bottom (SKA) plot.
\label{fig:spin-e0.1}}
\end{figure}

In Fig.~\ref{fig:spin-e0.1} we consider the spin measurement in a
PSR-SBH system with a MSP. The observations were based on
the same weekly scheme as in Fig.~\ref{fig:mass-e0.1}, but with
baseline of 10\,yr for a 100-m dish sensitivity (top plot) and 5\,yr
for the future telescope level (middle plot for the FAST, bottom
plot for the SKA). Clearly, the timing precision with a 100-m dish
would allow a black hole spin measurement only when the orbit is
compact enough ($P_{\rm b} \lesssim 0.5$\,days). The observations
conducted by the future telescopes, on the contrary, would achieve
the measurement with precision of order $\sim1$\% in only five years
for systems with wide ranges in both black hole mass and spin. Note
that here we use a low eccentricity of $e=0.1$ for the orbit. If the
pulsar is found to be a more eccentric binary, the spin is even
expected to be measurable for systems of orbital period up to
$\sim10$\,days.

%------------------------------------------------------------------------------%

\subsection{Quadrupole moment and no-hair theorem}
\label{ssec:qeff}

The quadrupole moment of the black hole induces a secular precession
of the pulsar orbit which is a few orders of magnitude smaller than
the spin contribution and thus is unlikely to be separable from the
overall precession.
\citep{wk99}. Fortunately, the pulsar orbit also endures periodic
perturbation by the quadrupolar field of the black hole. As will be
shown in this subsection, once the spin magnitude and orientation
have been determined by the overall precession of the orbit, in
principle this perturbation can be modelled in pulsar timing which
might lead to a determination of the black hole quadrupole moment.

\subsubsection{Modulation of the orbital motion and extraction of quadrupole} \label{sssec:qperturb}

The influence of the black hole quadrupole on pulsar's motion leads
to a variation in the R\"omer delay, which can be described by a
change in the coordinate position of the pulsar ${\bf r}'$,
described by %%
\begin{equation} \label{eq:qmotion}
  {\bf r}' = \left(r + \delta r^{(q)}\right)
             \left(\hat{\bf n} + \delta\hat{\bf n}^{(q)}\right) \;.
\end{equation}
The $\delta$-quantities can be derived by following \cite{g58} and
\cite{g59}, with slight modifications that account for the
dominating precession of the periastron caused by the mass monopole
\citep[for more details, see][]{lwk+12}. The variation scale is
proportional to a small dimensionless quantity $\epsilon$ which is
linked to the black hole quadrupole $Q_\bullet$ by: $\epsilon \equiv
-3Q_\bullet/a^2(1-e^2)^2$ \citep{g58,tpm86}. Here $a$ is the orbital
semi-major axis. The full expression of the R\"omer delay can then be
expanded with respect to $\epsilon$ as: %%
\begin{equation} \label{eq:qmodel}
  \Delta_{\rm R} = \Delta^{(0)}_{\rm R} +
    \delta\Delta_{\rm R} + \mathcal{O}(\epsilon^2),
\end{equation}
where $\delta\Delta_{\rm R}$ is of order $\epsilon$. Based on this
approximation as well as the MSS model, a new timing model has been
developed that includes the contribution of the black hole
quadrupole to first order in $\epsilon$ \citep{liu12}.

\subsubsection{Mock data analysis} \label{sssec:q simu}

As first suggested in \cite{wk99}, in a PSR-SBH system measurement of the
black hole quadrupole moment may not be possible since the amplitude
of the quadrupolar signal in one orbital period is only of order
$1\sim10$\,ns even for a highly relativistic system (e.g., $P_{\rm
b}\sim0.1$\,day, $e=0.9$). Nevertheless, due to the precession of
the orbit the quadrupolar feature will evolve on timescales of
years, which can increase the chance to detect the signal. To
investigate the circumstances where measurement of the black hole
quadrupole moment may become possible, we have performed extensive
mock data simulations. We assume weekly, four hour timing observations
of a MSP, for a
period of 20 yr with the sensitivity of the SKA. We extend our TOA
calculation and use the timing model as described in in
Section~\ref{sssec:qperturb} to account for the orbital periodic
effects due to the black hole quadrupole moment.

\begin{figure}
\centering
\includegraphics[scale=0.6]{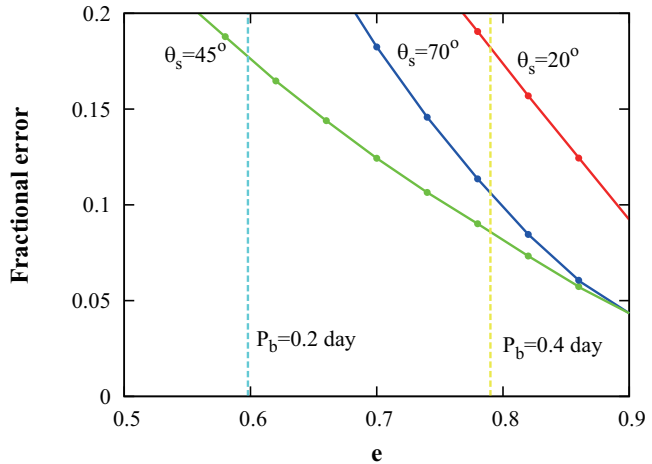}
\caption{Measurability of the quadrupole moment as function of orbital
eccentricity for PSR-SBH systems with a
30\,$M_{\odot}$ black hole, a merger time of 10\,Myr, and different spin inclination angles
$\theta_S$. Here we assume 20 years of observations of a MSP with the SKA
based on the weekly observing scheme mentioned in
Fig.~\ref{fig:mass-e0.1}. The eccentricity is varied from 0.5 to 0.9
which corresponds to an orbital period range of 0.16 and 0.88\,day.
The other system parameters are the same as in
Fig.~\ref{fig:spin-e0.1}. \label{fig:everr}}
\end{figure}

As indicated in \cite{wex98} and \cite{wk99}, in high eccentricity
orbits the quadrupole moment results in strong and sharp features in
timing residuals near the orbital periastron. The existence of such
features in the timing residuals would benefit the measurement of the
quadrupole moment, as
shown in Fig.~\ref{fig:everr}. Here we assume systems of a MSP with
a 30\,$M_{\odot}$ black hole, binary merging time of 10\,Myr, and
three different spin inclinations with respect to the line-of-sight
Clearly, the quadrupole moment is measurable only for systems of
high eccentricity ($e\gtrsim0.8$) and favorable geometry (e.g.,
$20^{\circ}<\theta_S<70^\circ$).

\begin{figure}
\centering
\includegraphics[scale=0.6]{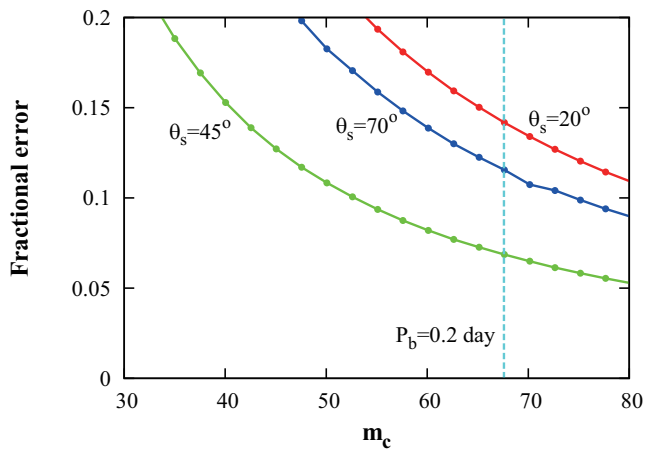}
\caption{Measurability of the quadrupole moment as function of black
hole mass for a 10\,Myr lifetime and mildly eccentric ($e = 0.5$)
PSR-SBH system, with different spin inclination angles $\theta_S$.
Note that the corresponding range of $P_{\rm b}$ is from 0.16\,days
(left) to 0.21\,days (right). The other system parameters are the
same as in Fig.~\ref{fig:spin-e0.1}. Here again, we assume 20\,yr
observations of a MSP with the SKA based on the weekly observing
scheme mentioned in Fig.~\ref{fig:mass-e0.1}. \label{fig:m2verr}}
\end{figure}

The mass of a SBH is usually found to be within the range of $5 \sim
30\,M_{\odot}$ \citep{zio08,sf08}. Recent studies have showen
that stars with very low metallicity can form stellar-mass black
holes with mass up to $80\,M_\odot$ from direct collapse
\citep{bbf+10}. Those high-mass stellar mass black holes are most
likely to be found in regions of very metal-poor environments, such
as globular clusters where frequent dynamic captures and 3-body
interactions due to the high stellar density might also allow
formation of a MSP-SBH system. Note that the quadrupolar field is
proportional to the cube of the black hole mass, therefore PSR-SBH
systems with a high-mass black hole would undoubtedly benefit the
measurement of quadrupole moment. Consequently, for simulations in
Fig.~\ref{fig:m2verr}, while using $e = 0.5$, $T_{\rm merge} =
10$\,Myr and three different spin inclinations, we extend the
parameter space of $m_\bullet$ to 80, the upper bound given by the
current formation studies \citep{bbf+10}. The results show that only
when $m_\bullet \gtrsim 70$ and there is a favorable geometry
(e.g., $20^\circ < \theta_S < 70^\circ$), there is a chance to measure the
quadrupole moment of the BH. Therefore, the measurement of the quadrupole moment
is possible if one finds a pulsar in orbit with a intermediate mass
black hole (IMBH), where $m_\bullet \sim 10^2$ to $10^4$, for instance in the
center of a globular cluster.
While there are some promising candidates for IMBH their existence
is still a matter of debate \citep{nm13}.

%%%%%%%%%%%%%%%%%%%%%%%%%%%%%%%%%%%%%%%%%%%%%%%%%%%%%%%%%%%%%%%%%%%%%%%%%%%%%%%%

\section{Testing scalar-tensor gravity with pulsar-black hole systems}
\label{sec:stg}

In the previous section we have discussed in detail, how timing a
pulsar in orbit with a stellar mass black hole can be used to probe
the properties of a black-hole spacetime. Questions like ``Is the
frame dragging in agreement with a Kerr solution, i.e.~$\chi \le
1$\,?'' and ``Does the quadrupole moment obey the relation given by
Eq.~(\ref{eq:q})\,?'' lie at the heart of these experiments. These
tests of GR's cosmic censorship conjecture and no-hair theorem with
a PSR-SBH system, if in agreement with GR, will at the same time
provide constraints on alternative theories of gravity that do not
allow for the Kerr solution as the outer spacetime of astrophysical
black holes. But even for alternatives to GR that also have the Kerr
metric as a solution for rotating black holes, e.g., the
scalar-tensor theories of \cite{de92,de93}, a PSR-SBH system would still be
an effective test-bed, especially with the sensitivity provided by
the next generation of radio telescopes. Like in
\cite{wle+13}\footnote{ Here we will present the details of the
calculation, which we could not in \cite{wle+13} because of the page
limit, and use a more realistic equation-of-state. }, we will
demonstrate this within the class of quadratic mono-scalar-tensor
theories, as introduced by \cite{de93,de96}, where the gravitational
interaction is mediated by a symmetric rank-2 tensor field
$g_{\mu\nu}$ and a scalar field $\varphi$. In the Einstein frame,
the field equations read %%
\begin{eqnarray}
   && R_{\mu\nu}^\ast = \frac{8\pi G_\ast}{c^4} \left(T_{\mu\nu}^\ast
     - \frac{1}{2} T^\ast g_{\mu\nu}^\ast \right)
     + 2 \partial_\mu \varphi \partial_\nu \varphi \;,\\
   && g_\ast^{\mu\nu} \nabla_\mu^\ast \nabla_\nu^\ast \varphi =
    -\frac{4\pi G_\ast}{c^4} \, \alpha(\varphi) \, T^\ast \;,
   \label{eq:boxphi}
\end{eqnarray}
where $g_{\mu\nu}^\ast$, $R_{\mu\nu}^\ast$, $T_{\mu\nu}^\ast$ and
$T^\ast$ are the metric tensor, the Ricci tensor, the stress-energy
tensor and the trace of the stress-energy tensor, respectively,
expressed in the Einstein frame (indicated by $\ast$). The constant
$G_\ast$ denotes the bare gravitational constant. The physical
(Jordan frame) metric $g_{\mu\nu}$ is conformally related to the
metric of the Einstein frame $g_{\mu\nu}^\ast$ by %%
\begin{equation}
  g_{\mu\nu} = e^{2 a(\varphi)} \, g_{\mu\nu}^\ast \;,
\end{equation}
and the coupling strength $\alpha(\varphi)$ between the scalar field
and matter (see Eq.~\ref{eq:boxphi}) is calculated according to %%
\begin{equation}
  \alpha(\varphi) = \partial a(\varphi) / \partial\varphi \;.
\end{equation}
In the class of quadratic mono-scalar-tensor theories of
\cite{de93,de96}, denoted as $T_1(\alpha_0,\beta_0)$ by
\cite{dam09}, one has %%
\begin{equation}
  a(\varphi) = \alpha_0 (\varphi - \varphi_0) +
               \frac{1}{2} \beta_0 (\varphi - \varphi_0)^2
\end{equation}
and %%
\begin{equation}
  \alpha(\varphi) = \alpha_0 + \beta_0(\varphi - \varphi_0) \;,
\end{equation}
with the two fundamental constants $\alpha_0$ and $\beta_0$. The
quantity $\varphi_0$ denotes the asymptotic value of $\varphi$ at
spatial infinity. Without loss of generality, here we set $\varphi_0
= 0$. In the weak-field limit ($\varphi \simeq \varphi_0$),
$\alpha_0$ and $\beta_0$ give the strength of the linear and
quadratic coupling of $\varphi$ to matter respectively.
Jordan-Fierz-Brans-Dicke gravity is equivalent to the special case
$\beta_0 \equiv 0$, and the Brans-Dicke parameter is given by %%
\begin{equation}
  \omega_{\rm BD} = \frac{1}{2}\left(\frac{1}{\alpha_0^2} - 3 \right) \;.
\end{equation}

For a strongly self-gravitating body, like a pulsar, the coupling
$\alpha_0$ has to be replaced by an effective (body-dependent)
coupling strength, which is defined as %%
\begin{equation}
  \alpha_{\rm p} \equiv
    \frac{\partial\ln M_{\rm p}}{\partial\varphi_a} \;,
\end{equation}
where $\varphi_a$ is the asymptotic field felt by the pulsar.
Furthermore, in binary pulsar experiments one needs the scalar-field
derivative of the effective coupling strength, i.e. %%
\begin{equation}
  \beta_{\rm p} \equiv \frac{\partial\alpha_{\rm p}}{\partial\varphi_a} \;.
\end{equation}
The two derivatives above have to be taken for fixed values of the
baryonic mass of the pulsar.

Due to their large asymmetry in compactness, binaries of a pulsar
and a white dwarf ($\alpha_{\rm WD} \simeq \alpha_0$) are
particularly interesting for constraining scalar-tensor theories of
gravity, as theses systems should lose orbital energy at a much
higher rate due to the emission of dipolar gravitational waves
\citep{will93,de96}. A combination of such experiments along with
Solar-System tests has already placed tight constraints on the
$(\alpha_0,\beta_0)$ parameter space (see Fig.~7 in \cite{fwe+12}).
In brief, the area $|\alpha_0| > 0.003$ and $\beta_0 < -4.5$ is
already excluded. The discovery of a massive pulsar ($m_{\rm p}
\simeq 2.0\,M_\odot$) in a relativistic orbit added further
restrictions, in particular giving $\beta_0 \gtrsim -4.3$
\citep{afw+13}. However, as already pointed out by \cite{de98} a
PSR-SBH system would generally be even more asymmetric, since the
no-scalar-hair theorem for black holes in scalar-tensor gravity
gives \citep{haw72,de92}
\begin{equation} \label{eq:aAbABH}
  \alpha_\bullet = 0 \;, \quad \beta_\bullet = 0 \;.
\end{equation}
Strictly speaking, this is only valid for stationary black holes
where the metric is asymptotically flat and the scalar field is
asymptotically constant (see \cite{bcg+13} for a detailed discussion
on the validity of the classical no-hair theorem and a generalized
no-hair theorem in scalar tensor gravity). However, as we will
discuss below, this is still a very good approximation in PSR-SBH
systems.

In this section we will demonstrate that the discovery and timing of
a PSR-SBH system in a $P_{\rm b} \lesssim 5$\,days orbit has the
potential to significantly improve existing constraints on
scalar-tensor gravity, especially with the next generation of radio
telescopes.

%------------------------------------------------------------------------------%

\subsection{Post-Keplerian parameters in PSR-SBH systems}
\label{ssec:pkpsrbh}

The parametrized post-Keplerian formalism can be used not only in
GR, but also in a wide class of alternative theories of gravity,
including the $T_1(\alpha_0,\beta_0)$ class of scalar-tensor
theories discussed here \citep{dam88,dt92}. The explicit expressions
for the post-Keplerian parameters in $T_1(\alpha_0,\beta_0)$ can be
found in \cite{de96}. For a PSR-SBH system Eq.~(\ref{eq:aAbABH})
considerably simplify these expressions. For the quasi-stationary PK
parameters one finds %%
\begin{eqnarray}
  \dot{\omega} &=& \frac{3}{1-e^2} \,
    \left(\frac{P_{\rm b}}{2\pi}\right)^{-5/3}
    \frac{G_\ast^{2/3}}{c^2} \, (M_{\rm p} + M_\bullet)^{2/3} \;,
    \label{eq:PKomdotSTG} \\
  \gamma &=& e\left(\frac{P_{\rm b}}{2\pi}\right)^{1/3}
    \frac{G_\ast^{2/3}}{c^2} \,
    \frac{(M_{\rm p} + 2 M_\bullet) M_\bullet}
         {(M_{\rm p} + M_\bullet)^{4/3}} \;, \\
  s_{\rm Sh} &=& x\left(\frac{P_{\rm b}}{2\pi}\right)^{-2/3}
    c \, G_\ast^{-1/3} \,
    \frac{(M_{\rm p} + M_\bullet)^{2/3}}{M_\bullet} \;,
    \label{eq:PKsSTG}\\
  r_{\rm Sh} &=& \frac{G_\ast}{c^3} \, M_\bullet \;.
    \label{eq:PKrSTG}
\end{eqnarray}
The bare gravitational constant $G_\ast$ is related to the
gravitational constant measured in a Cavendish experiment by $G =
G_\ast(1 + \alpha_0^2)$. Consequently, modulo an unobservable
rescaling of the masses\footnote{It is interesting to point out, that
from Cassini we already know that $\alpha_0^2 \lesssim 10^{-5}$.
Consequently, a rescaling of the masses by $(1 + \alpha_0^2)^{-1}$
is anyhow (generally) small compared to the implicite rescaling due
to the unknown systemic radial velocity \citep{dt92}.}, the PK
parameters given above are identical to the quasi-stationary PK
parameters in GR (cf.\
Eqs.~\ref{eq:PKomdot},\ref{eq:PKgamma},\ref{eq:PKshap},\ref{eq:PKs}).
This agrees with the findings of \cite{mw13}, that through first
post-Newtonian order ($c^{-2}$), the motion of a black-hole
neutron-star system in scalar-tensor gravity is identical to that in
general relativity.

However, the situation is different for the gravitational wave
damping (PK parameter $\dot{P}_{\rm b}$). Due to the scalar charge
of the neutron star, the system would emit gravitational radiation
of all multipoles, since the system can now loose energy to scalar
waves in addition to energy loss into tensor waves
\citep{will93,de92}. The scalar monopole and quadrupole
contributions enter the orbital dynamics at the 2.5 post-Newtonian
(PN) level (order $c^{-5}$), and are normally much smaller than the
tensor-quadrupole contribution, given the existing constraints on
$\alpha_p$ for pulsars up to $2\,M_\odot$ \citep{fwe+12,afw+13}.
Concerning the scalar dipole radiation, the situation is very
different, as this contribution affects the orbital dynamics already
at the 1.5 post-Newtonian level (order $c^{-3}$) and is therefore
enhanced (compared to the tensor waves) by a large factor of
$c^2/v^2$, where $v$ is the velocity of the relative motion of the
binary. The orbital averaged change in the orbital period up to 2.5PN
order can be written as a sum of scalar ($\varphi^\ast$) and tensor
($g^\ast$) contributions, %%
\begin{eqnarray}\label{eq:PbdotSum}
  \dot{P}_{\rm b} &=&
  \dot{P}_{\rm b,\varphi^\ast}^{\rm Monopole, 2.5PN} +
  \nonumber\\&&
  \dot{P}_{\rm b,\varphi^\ast}^{\rm Dipole, 1.5PN} +
  \dot{P}_{\rm b,\varphi^\ast}^{\rm Dipole, 2.5PN} +
  \nonumber\\&&
  \dot{P}_{\rm b,\varphi^\ast}^{\rm Quadrupole, 2.5PN} +
  \nonumber\\&&
  \dot{P}_{\rm b,g^\ast}^{\rm Quadrupole, 2.5PN}
\end{eqnarray}
where for a PSR-BH system ($\alpha_\bullet = \beta_\bullet=0$) all
these terms depend on $P_b$, $e$, the (gravitational) masses $G_\ast
M_{\rm p}$ and $G_\ast M_\bullet$, and all $\varphi^\ast$-terms are
proportional to $\alpha_{\rm p}^2$ \citep{de92}. To leading order we
have
\begin{eqnarray}
  \dot{P}_{\rm b}
  &=& -\frac{192\pi}{5} \left(\frac{P_{\rm b}}{2\pi}\right)^{-5/3}
       \frac{G_\ast^{5/3}}{c^5} \,
       \frac{M_{\rm p} M_\bullet}{(M_{\rm p} + M_\bullet)^{1/3}} \, f(e)
     \nonumber\\
  & & -\frac{4\pi^2}{P_{\rm b}} \,
       \frac{G_\ast}{c^3} \,
       \frac{M_{\rm p} M_\bullet}{M_{\rm p} + M_\bullet} \,
       \frac{1 + e^2/2}{(1-e^2)^{5/2}} \,
       \alpha^2_{\rm p} \;.
  \label{eq:PKpbdotSTG}
\end{eqnarray}
For the still allowed region of $T_1(\alpha_0,\beta_0)$ one can show
that the scalar 2.5PN monopole, dipole and quadrupole terms in
Eq.~(\ref{eq:PbdotSum}) are several orders of magnitude smaller than
the terms in Eq.~(\ref{eq:PKpbdotSTG}). The last term (proportional
to $\alpha_p^2$) is the key to constrain scalar-tensor theories with
PSR-SBH systems, while the other PK parameters
(Eqs.~\ref{eq:PKomdotSTG} to \ref{eq:PKrSTG}) merely provide the
constraints for the, a priori unknown, masses $M_{\rm p}$ and
$M_\bullet$. For a given EoS, $\alpha_{\rm p}$ is a function of
$\alpha_0$, $\beta_0$ and $M_{\rm p}$. Based on this, a measurement
or constraint of $\dot{P}_{\rm b}$ in a PSR-SBH system, in
combination with the measurement of at least two more PK parameters,
can be converted into an exclusion area within the
$\alpha_0$-$\beta_0$ plane of $T_1(\alpha_0,\beta_0)$ theories.

As discussed in Section~\ref{sssec:fd}, the observed $\dot{\omega}$
can be significantly influenced by the frame-dragging caused by the
rotation of the black hole. For simplicity, we assume here that the
rotation of the black hole is sufficiently small. In practice, if
the companion is a fast rotating black hole, one would need a
self-consistent analysis, that at the same time provides the spin of
the black hole from the Lense-Thirring precession of the orbit.
Alternatively, depending on the orientation and eccentricity of the
system, the Shapiro delay might give an independent access to
$M_{\rm p}$ and $M_\bullet$.

Finally, the scalar no-hair theorem can be shown to be valid in
PSR-SBH systems, based on an order of magnitude estimation. If a
pulsar that orbits a black hole in an eccentric orbit carries a
scalar charge, it induces a time-variant scalar field at the
location of the black hole, given by %%
\begin{equation}
  \varphi_r^{(\rm p)} \approx \varphi_0 + \frac{G M_{\rm p}}{c^2 r}
     \alpha_{\rm p} \;,
\end{equation}
where $r$ is the relative separation between pulsar and black hole.
For an eccentric orbit ($e \ne 0$), $r$ changes on a timescale of
the orbital period $P_{\rm b}$. According to Jacobson's ``Miracle
Hair Growth Formula'' \citep{jac99,bcg+13} this induces a scalar
charge on a black hole. The corresponding (effective) scalar
coupling can be estimated by %%
\begin{equation}
  \alpha_\bullet^{\rm induced} \approx
   4 \frac{G M_\bullet}{c^3} \, \partial_t \varphi_r^{(\rm p)}
  \approx 4 \frac{G M_\bullet}{c^2 r} \,
            \frac{G M_{\rm p}}{c^2 r} \,
            \frac{\dot{r}}{c} \,
            \alpha_{\rm p} \;.
\end{equation}
For a compact PSR-SBH system ($r \sim 1$\,lt-s) one finds $G
M_\bullet/c^2 r \sim 10^{-5}$, $G M_{\rm p}/c^2 r \sim 10^{-6}$, and
$\dot{r}/c \sim 10^{-3}$. Consequently, the induced scalar coupling
is absolutely negligible for the gravity tests outlined in this
section.

%------------------------------------------------------------------------------%

\subsection{Mock data analysis}
\label{ssec:TS simu}

The potential of a PSR-SBH system in constraining scalar-tensor
theories of gravity can be demonstrated by mock-data simulations.
Here we assume a PSR-SBH system with $P_{\rm b}=5$\,day, $e=0.8$,
and $m_{\rm p}=1.4$. A stiff EoS allowing a maximum neutron star
mass of $\sim 2.5\,M_\odot$ \citep{mpa87}, is used to calculate
$\alpha_{\rm p}$ for any given theory $T_1(\alpha_0,\beta_0)$. Note
that this choice of EoS leads to conservative constrains, as the
application of a softer EoS would generally lead to larger scalar
charges for the neutron star and therefore more stringent limits.
For the black hole, we have assumed a non-rotating ($\chi = 0$)
10\,$M_\odot$ black hole, because a significant spin would only
complicate the timing analysis (see Section~\ref{sec:GR_test}) but
should not greatly influence the constraints on scalar-tensor
theories. The applied observational scheme is retained as in
Section~\ref{sssec:masssimu}. The simulated TOAs are based on the
orbital dynamics of GR, and fitted with the DD timing model to
estimate the PK parameters. These measurements are then confronted
with Eqs.~(\ref{eq:PKomdotSTG}) to (\ref{eq:PKpbdotSTG}), in order
to exclude those areas of the $\alpha_0$-$ \beta_0$ for which no
pair $(M_{\rm p},M_\bullet)$ can be found such that all calculated
PK parameters agree with the fitted values within the measurement
errors. A more detailed description of tests of
$T_1(\alpha_0,\beta_0)$ theories with PK parameters can be found in
\cite{de96}.

\begin{figure}
\centering
\includegraphics[scale=0.31]{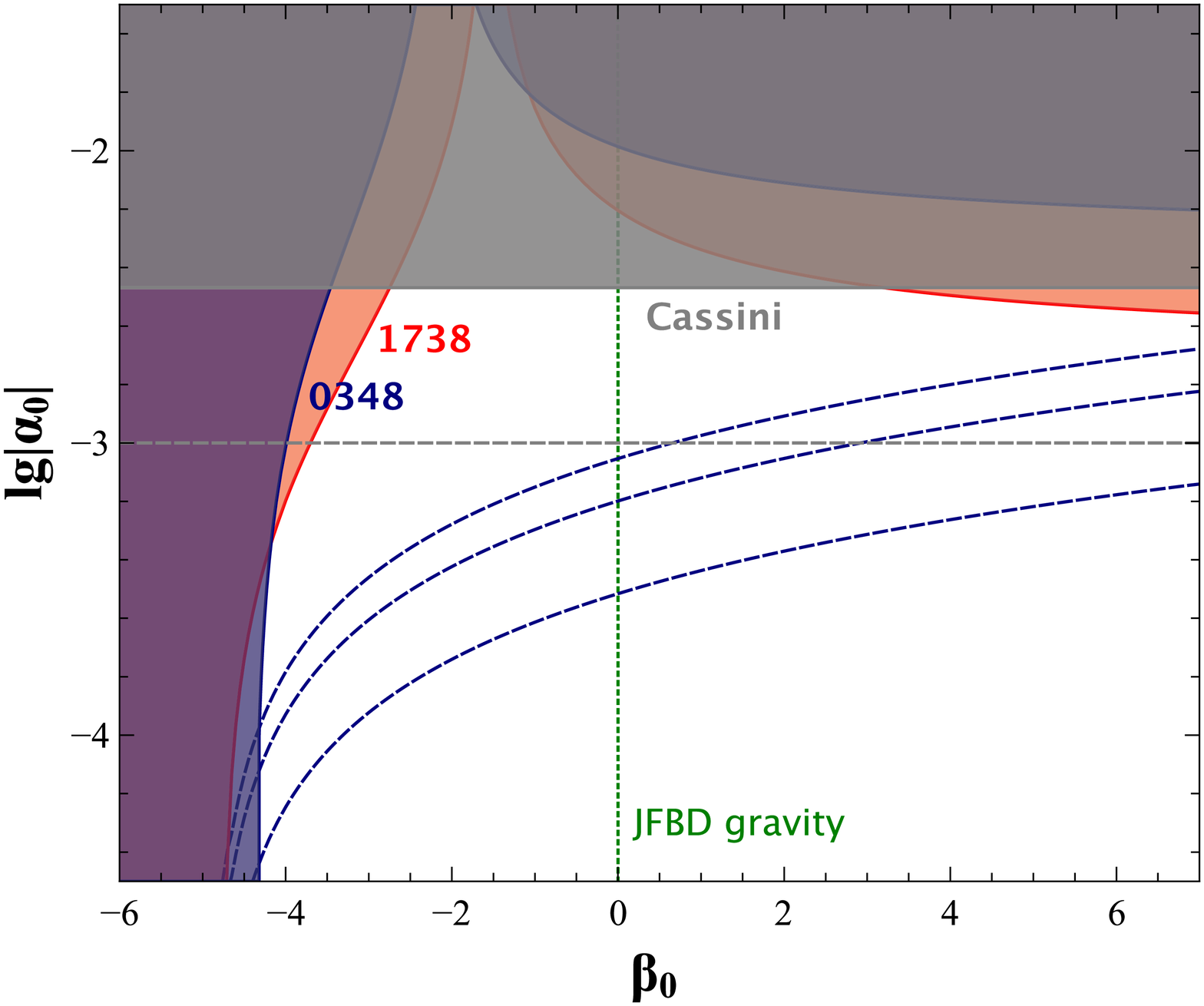}
\caption{Constraints on $T_1(\alpha_0,\beta_0)$ scalar tensor
theories. Exclusion areas are based on current Solar system and
pulsar experiments, taken from \citet{bit03} ('Cassini'),
\citet{fwe+12} ('1738'), and \citet{afw+13} ('0348'). Blue dashed lines
are based on simulations for a MSP-SBH system: from top to bottom:
10 years with a 100-m class, 5 years with the FAST and SKA (details
can be found in the text). The vertical dashed green line at
$\beta_0 = 0$ indicates Jordan-Fierz-Brans-Dicke gravity. The horizontal
dashed grey line indicates the limit expected from near future Solar-system
experiments, foremost the astrometric satellite GAIA.
\label{fig:TS-e0.8}}
\end{figure}

Fig.~\ref{fig:TS-e0.8} presents the results for three different
scenarios: 10 years timing with a 100-m class, 5 years with the FAST
and SKA. The simulations show clearly, that a PSR-SBH system would
be a great test-bed for scalar-tensor gravity, in particular with
the FAST and the SKA. A few years timing observations of a MSP-SBH
system with these future radio telescopes would lead to
significantly better constraints, with the potential to greatly
exceed Solar-System experiments for all values of $\beta_0$
(including Jordan-Fierz-Brans-Dicke gravity $\beta_0 = 0$), similar
to the expectation from GAIA.

As a final remark, for very large $\beta_0$ the neutron star nearly
completely de-scalarizes, making a PSR-BH system a less sensitive
test for scalar-tensor gravity. For instance, for a 1.4\,$M_\odot$
neutron-star and $\beta_0 = 20$ one finds $\alpha_p \simeq
0.14\,\alpha_0$, and for $\beta_0 = 200$ one gets $\alpha_p \simeq
0.02\,\alpha_0$.

%%%%%%%%%%%%%%%%%%%%%%%%%%%%%%%%%%%%%%%%%%%%%%%%%%%%%%%%%%%%%%%%%%%%%%%%%%%%%%%%

\section{External factors}
\label{sec:dis}

In practice, there are other effects that can contaminate the
measured properties used for the GR tests enabled by PSR-SBH
systems. The observed secular change of the orbital projected
semi-major axis can be influenced by shrinking of the orbit due to
gravitational radiation, proper motion of mass centre
\citep{kop96}, geodetic precession of the pulsar's spin \citep{dt92},
and a
varying Doppler shift caused by secular change in the distance of
the binary mass centre and the relative acceleration in the gravitational field
of the Galaxy \citep{shk70,dt91}. Assuming a 1.4\,$M_\odot$
pulsar, one can obtain the comparison of the gravitational wave
damping contribution with the spin-orbit effect in GR as \citep{pet64}:
\begin{equation}
  \left|\frac{\dot{x}_{\rm gw}}{\dot{x}_{\rm s}}\right| \simeq
    4.8\times10^{-6} \,\frac{f(e)}{|\chi\sin\theta_S\sin\Phi\cot i|}
   \left(\frac{P_{\rm b}}{\rm
   1~day}\right)^{-2/3}m_{\bullet}^{-1/3}.
\label{eq:xdot_s v xdot_gw}
\end{equation}
For a black hole of $m_{\bullet}=10$, $\chi=0.9$, and an orbit of
$P_{\rm b}=0.1$\,day, $e=0.8$, $\theta_S=\Phi=i=45^{\circ}$, the
ratio appears to be $\sim10^{-4}$, suggesting that normally the
spin-orbit coupling dominates the secular change of $x$ over
gravitational wave damping. The contribution by proper motion of the mass
centre could be significant if the binary is close to the solar
system, but in this case can also be measured and subtracted by high
precision astrometry \citep[e.g.][]{dbl+13} or pulsar timing itself.
The contribution by the pulsar's geodetic precession, following
\cite{dt92}, can be compared with $\dot{x}_{\rm s}$ by
\begin{equation}
\left|\frac{\dot{x}_{\rm geo}}{\dot{x}_{\rm s}}
  \right|\sim6\times10^{-3}\,\chi^{-1}\left(\frac{P}{\rm
1\,s}\right)\left(\frac{P_{\rm b}}{\rm
1\,day}\right)^{-2/3}\left(\frac{m_{\bullet}}{10}\right)^{-1/3},
\end{equation}
where $P$ is the pulsar rotational period. Therefore, the effect is
stronger for PSR-BH system with a slow pulsar. With $P=0.5$\,s,
$P_{\rm b}=0.2$\,day, $m_{\bullet}=10$, and $\chi=0.9$, the ratio
$\sim7\times10^{-3}$, indicating that this effect needs to be taken
into account only for a slow pulsar in a very compact orbit. The
effect of varying Doppler shift contains contributions from relative
line-of-sight Galactic acceleration and traverse velocity (the
Shklovskii term) between the system and the solar system barycentre.
Its contribution to $\dot{x}$, as argued in \citep{wk99}, will not
be important unless the binary system is in the central region of
our galaxy.

The observed secular change of orbital period can also be
contaminated by the varying Doppler shift. Recent studies have shown
that based on the current Galactic potential model, the contribution
to $\dot{P}_{\rm b}$ caused by the relative line-of-sight acceleration can be corrected at a level of
$10^{-15}$ to $10^{-16}$ \citep{lwj+09,fwe+12}. The
determination of the Shklovskii effect depends on the measurement
accuracies of proper motion and distance, and can be expected with
high precision from astrometry or pulsar timing itself
\citep[e.g.][]{stw+11,dbl+13}. In this case, the precision of the
Galactic potential model is more likely to be a limiting factor for
dipole radiation tests. The varying Doppler shift may also limit the
precision of mass determination with $\dot{P}_{\rm b}$ and thus the
cosmic censorship conjecture test, especially for wide orbits with
small $\dot{P}_{\rm b}$ ($\lesssim10^{-14}$). Nevertheless, one might
be able to use the measurement of the Shapiro delay instead to determine the
masses, which in this case would be anyway determined with better
precision for systems of sufficiently edge-on geometry (i.e.,
$i\gtrsim60^{\circ}$).

The geodetic precession of the pulsar spin will result variation in
pulse profiles \citep[e.g.][]{kra98}, which may complicate or limit
the pulsar timing precision. For the worst case, it might even turn
the radiation beam away from our line-of-sight \citep{pmk+10}. To
leading order the spin precession is given by the geodetic
precession rate, which for $m_{\bullet} \gg m_{\rm p}$ reads
\citep{bo75b}
\begin{equation}
\Omega_{\rm geod}\approx(0.5\,{\rm deg/yr})\frac{1}{1-e^2}
                         \left(\frac{P_{\rm b}}{1\,{\rm day}}\right)^{-5/3}\left(\frac{m_{\bullet}}{10}\right)^{2/3}.
\end{equation}
Consequently, for an orbit of $P_{\rm b}=0.1$\,days,
$m_{\bullet}=10$, and $e=0.1$ the precession rate is roughly
22\,deg/yr. Fortunately, the spin geometry of the pulsar can also be
studied from polarimetric information \citep[e.g.][]{kra98,mks+10},
which may provide a method to properly model the profile evolution
and still enable high precision timing. On the other hand, the
observation of the geodetic precession can give access to a further
post-Keplerian parameter, if the precession rate can be determined
independently. Unfortunately, even if an independent measurement of
the precession rate is possible, it is not expected to give a high
precision value, like for the other post-Keplerian parameters (cf.\
\cite{bkk+08} and \cite{fst14} for two binary pulsar systems, where
the geodetic precession rate has been measured). Still, there could
be important information coming from modeling the geodetic
precession of the pulsar. In general, the spin of the SBH companion
is much larger than that of the pulsar. But in the case of a fast
rotating pulsar (MSP) and a slowly rotating ($\chi \lesssim 0.1$)
low-mass ($m_\bullet \lesssim 5$) SBH, the pulsar spin can easily
reach $\sim$10\% of the SBH spin, giving a corresponding
contribution to the relativistic spin-orbit coupling. In such a case
the orientation of the pulsar spin, coming from the observation of
the geodetic precession, is needed to extract the SBH spin from the
observed Lense-Thirring precession of the orbit.

\section{Searches for pulsar--black hole binaries} \label{sec:search}
Searches for undiscovered pulsars have been extensively performed
ever since the first discovery in 1967 \citep{hbp+68}. Despite
improvements in both observational hardware and data processing
techniques, the discovery of a PSR-SBH system has so far eluded
pulsar astronomers. Although it is expected that PSR-SBH systems are
rare, it can be shown that observational selection effects could
have played an important role in the non-detection. Pulsars in
binary systems show periodic changes in their spin frequency because
of Doppler effects induced by orbital motion. In compact, high-mass,
or eccentric systems these effects can manifest themselves within
the timescale of individual pulsar survey observations i.e.\
typically of the order of minutes. Standard Fourier based searches,
that look for significant features in the power spectrum of
dedispersed time series, are not sensitive to pulsar spin
frequencies that change, since power is smeared over a number of
spectral bins, reducing the detection signal-to-noise (S/N) ratio
\citep[see e.g.][]{jk91, rem02}. To combat the detrimental effects
of orbital motion, pulsar searches typically employ computationally
intensive binary search algorithms, such as ``acceleration
searches'' \citep[for a summary of current methods, see][]{lk05}.
Like in pulsar `drift scan' surveys \citep[e.g.][]{dsm+13}, many of
these effects can be overcome by reducing the dwell time of survey
observations; a possibility offered by the supreme instantaneous
sensitivity of next generation telescopes. In this section we
investigate the basic search requirements for the detection of
PSR-SBH systems.

\subsection{Acceleration searches and computational considerations}
\label{ssec:acc_search}

Changes in the apparent spin frequency are caused by the varying
line-of-sight (l.o.s) velocity of the pulsar, $v(t)$, as it orbits
its companion. The apparent spin frequency as a function of time,
$\nu_{\rm app}(t)$, is given by the Doppler formula:
\begin{equation}\label{eq:doppler}
  \nu_{\rm app}(t) = \nu\left(1 - \frac{v(t)}{c}\right),
\end{equation}
where $v(t)$ can be described by five Keplerian orbital parameters.
If these parameters are known, it is possible to resample the time
series and transform it into a frame inertial with respect to the
pulsar. Standard Fourier methods can then be used to detect the
periodic signal from a pulsar with no reduction in S/N. However in a
blind search, where the orbital parameters are initially unknown, a
five dimensional search is computationally prohibitive \citep[see
e.g.][]{kek+13}. If the observing time ($T_{\rm obs}$) is a small
fraction of the orbital period, the l.o.s velocity can be
approximated in a Taylor expansion:
\begin{equation}\label{eq:taylor}
v(t)\approx v_{0} + a_{0}t + j_{0}t^2/2 + ...,
\end{equation}
where $v_{0}$, $a_{0}$, and $j_{0}$ are the average values of
velocity, acceleration, and rate of change of acceleration --
``jerk'', respectively. When $T_{\rm obs}$ is sufficiently small
\citep[see e.g.\ $T_{\rm obs} \lesssim P_{\rm b}/10$ from][ Ng et
al. in prep.]{rce03}, time or equivalent frequency domain searches
in only constant acceleration, given by the first order term, $v(t)
\simeq a_{0}t$ (the constant $v_0$ can be dropped here), are
effective \citep[e.g.][]{clf+00, rem02, ekl+13}.

For the detection of binary systems where large changes in
acceleration are observed (see Section~\ref{ssec:sims}),
improvements in S/N are offered by searches in higher order velocity
derivatives \citep[see e.g.][]{ekl+13,
  kek+13, blw13}, at the cost of at least an order of magnitude more
computational operations. It is expected that pulsar searches to be
performed with the SKA will be performed in real time or ``pseudo real
time''\footnote{\footnotesize To achieve the specified limiting
  sensitivity to periodic pulsed signals, the full integration time
  will be completed before data processing can commence.}, to cope
with data output volumes. Because first order acceleration searches
already constitute a significant data processing
task\footnote{\footnotesize In Table~24 of \cite{dtm+13}, it is shown
  that the data processing for a pulsar survey with SKA1-mid is
  dominated by the acceleration search processing load of nearly 10
  Peta operations per second; 10 times the processing power offered by
  the Einstein@Home network as of January 2013:
  http://einstein.phys.uwm.edu/} higher order realtime searches are
unlikely to be performed, unless adequate computational hardware or
software becomes available.\footnote{\footnotesize Following the
SKA1-mid survey parameters outlined in Table~24 of \cite{dtm+13},
and correcting for a jerk of $\pm\,0.2\,{\rm m\,s^{-3}}$ (a value
that could be observed in known compact double neutron star systems:
see Section~\ref{sssec:accn_char}), we find that a pseudo real time
acceleration and jerk search would require over an Exa Flop per
second of computation. By covering the increased acceleration and
jerk parameter ranges expected to be observed in PSR-SBH
(Section~\ref{sssec:accn_char}), the computational cost could be
higher.} With reference to the latter, we notice that significant
progress in speeding up various aspects of pulsar search code has
been made in recent years, primarily through the use of GPU
technology \citep[see e.g.][ Luo
2013\footnote{https://github.com/jintaoluo/presto\_on\_gpu}, Barr et
al. in
prep\footnote{https://github.com/ewanbarr/peasoup}]{bbf10,mks+11,akg+12,bbb+12}.
In the following sections our investigation is limited to searches
utilizing the constant acceleration approximation only.

\begin{figure}
\centering
\includegraphics[angle=-90,scale=0.35]{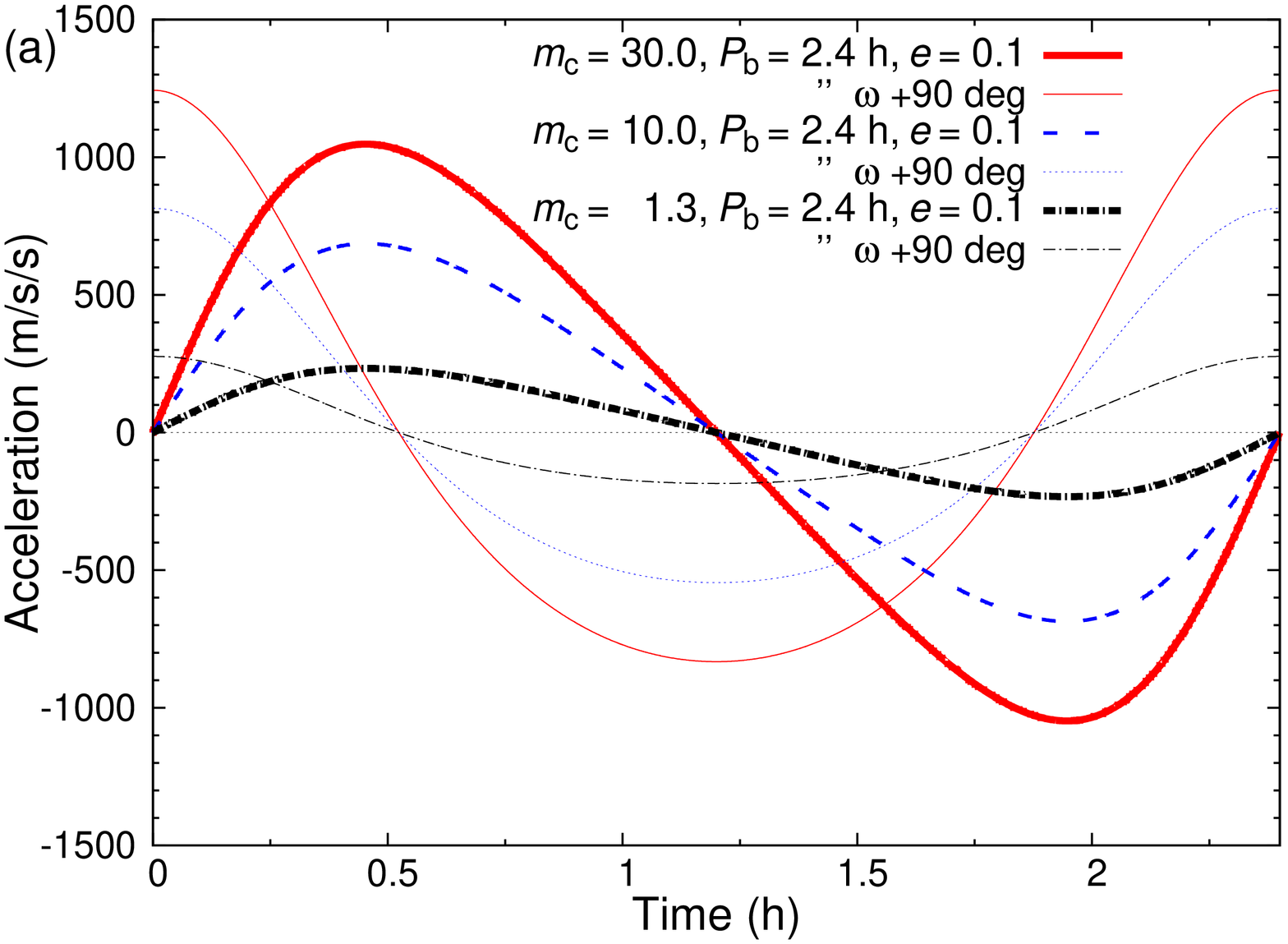}
\includegraphics[angle=-90,scale=0.35]{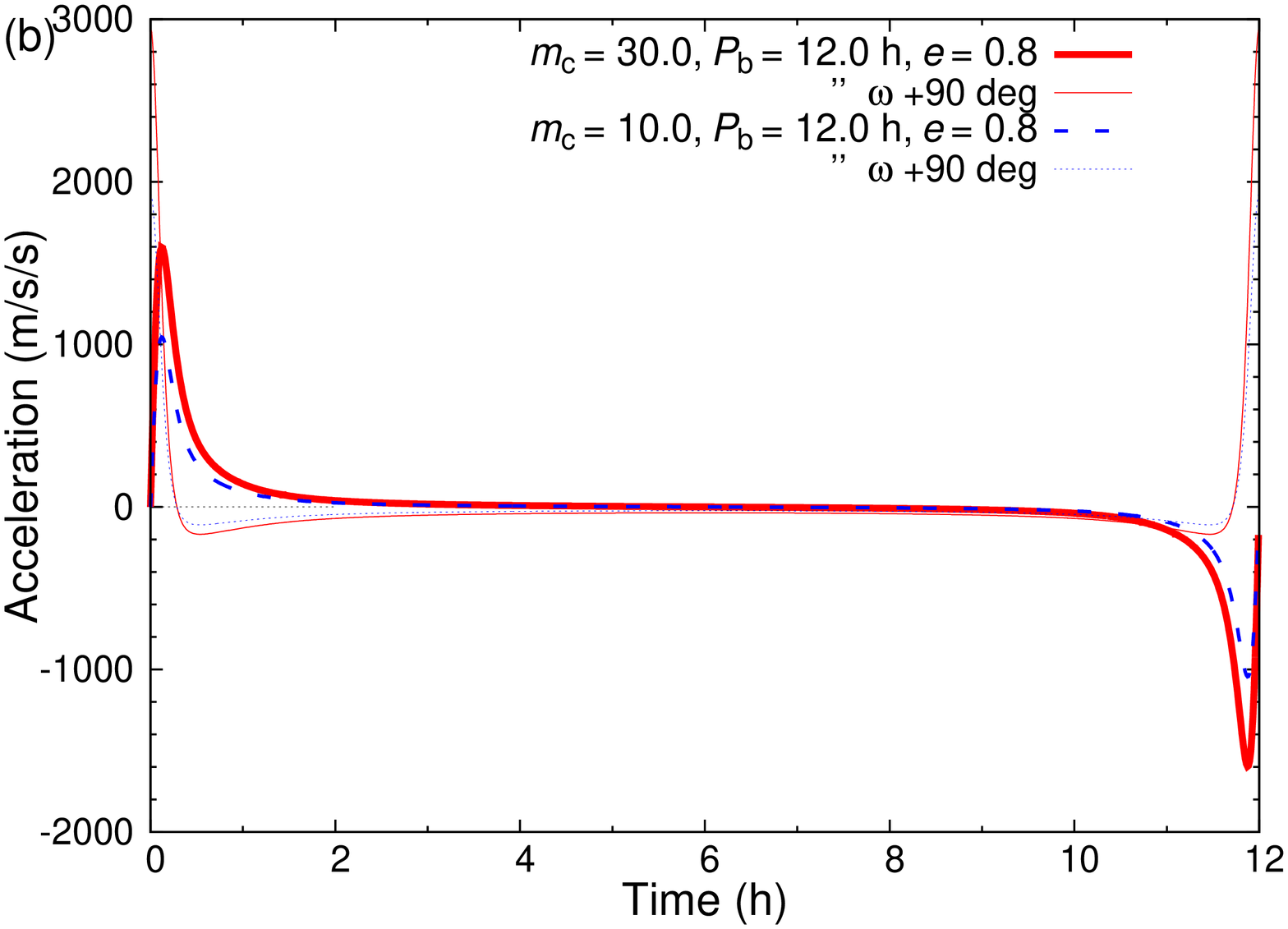}
\includegraphics[angle=-90,scale=0.35]{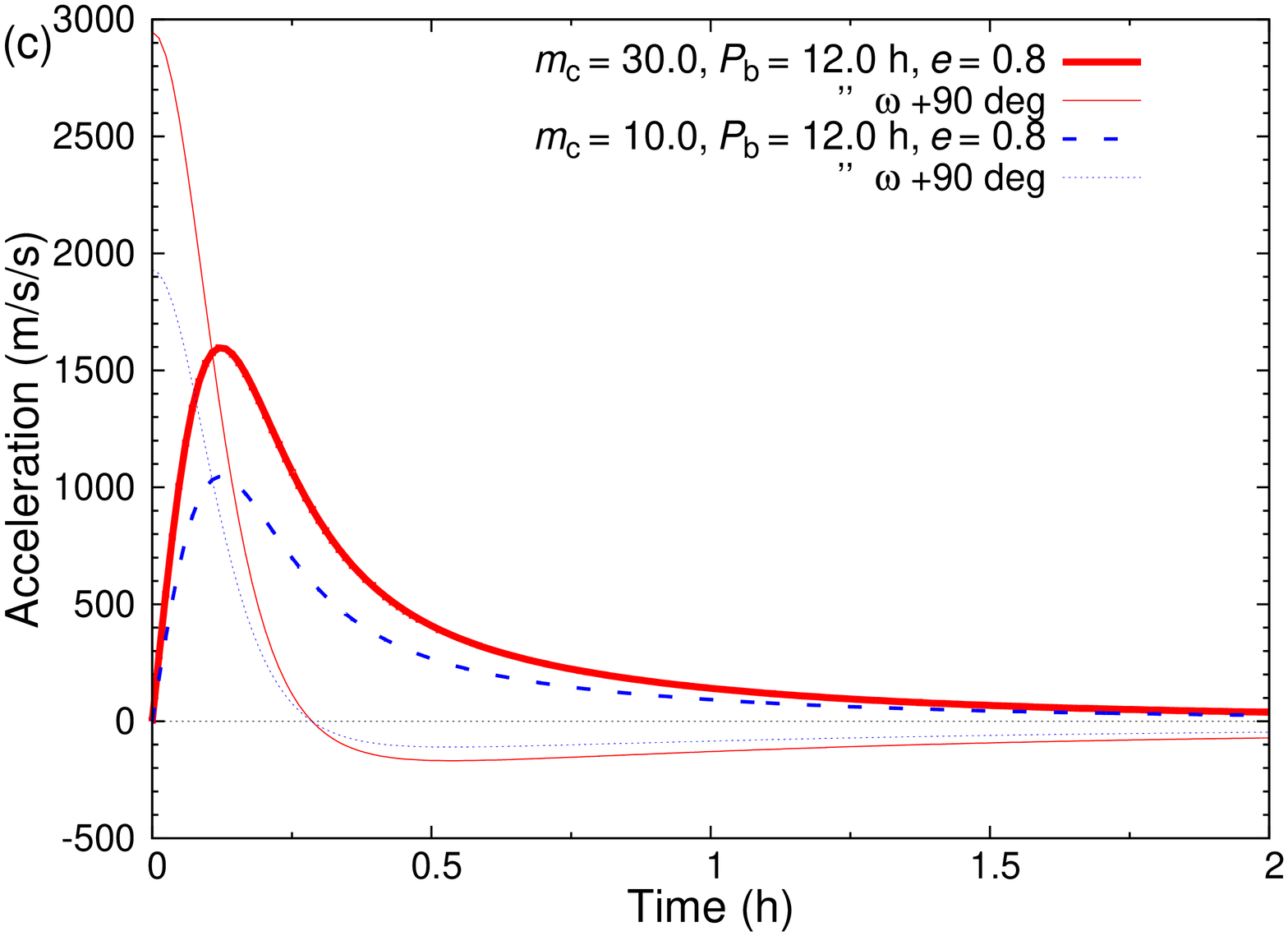}
\caption{The expected l.o.s acceleration as a function of time in
speculative relativistic binary pulsar systems. Panel~(a) shows
compact, $P_{\rm b}=2.4$ hours, near circular, $e=0.1$, systems
while panel~(b) shows longer orbital period, $P_{\rm b}=12.0$ hours,
eccentric, $e=0.8$, systems. Panel~(c) shows a zoom in of the first
2\,hr of the latter. In both cases black hole masses of
10\,$M_{\odot}$ and 30\,$M_{\odot}$, and extremes in longitude of
periastron, $\omega$ have been plotted. In panel~(a) the expected
values for a compact DNS system (pulsar companion mass of
1.3\,$M_{\odot}$) have also been included. In all cases the systems
are viewed edge-on i.e. $i=90^{\circ}$.\label{fig:psr_bh_accels}}
\end{figure}

\subsubsection{PSR-SBH orbital acceleration characteristics}\label{sssec:accn_char}
It is useful to consider the degree of orbital acceleration that
might be observable in compact or eccentric PSR-SBH systems as this
gives the parameter space that needs to be searched with
acceleration searches, and rough estimates of the corresponding
level of computation required (Section~\ref{sssec:comp_scale}). In
Fig.~\ref{fig:psr_bh_accels}, panels~(a), (b) and (c) the expected
value of l.o.s acceleration as a function of observation time is
given for various extreme PSR-SBH systems. Panel~(a)  shows the
results for compact ($P_{\rm b}=2.4$\,h) near circular systems
($e=0.1$) and panels~(b) and (c) show longer orbital period ($P_{\rm
  b}=12.0$\,h) eccentric ($e=0.8$) systems. In both cases the binary
systems are viewed edge-on ($i=90^{\circ}$), where the effects of
l.o.s. motion are strongest, and thick lines (both solid and dashed)
show the acceleration for systems with $\omega=0^\circ$ and thin solid and
dashed lines show systems which have the major axis of the pulsar
orbit pointed toward the l.o.s ($\omega=90^{\circ}$). L.o.s pulsar
acceleration values have been plotted for two companion black hole
masses of 10\,$M_{\odot}$ and 30\,$M_{\odot}$, and in panel~(a)
values for a neutron star companion of 1.3\,$M_{\odot}$ have also
been plotted.  The l.o.s pulsar acceleration values have all been
calculated using Eq.~(4) in \cite{fkl01,fkl09} which is derived from
Kepler's laws.

From Fig.~\ref{fig:psr_bh_accels} we would like to draw the readers
attention to two points. Firstly, and as expected, the maximum
accelerations that might be observable in PSR-SBH systems can
greatly exceed those seen in currently known highly relativistic
binary systems. This can be seen by comparison of the acceleration
values plotted for the 1.3\,$M_{\odot}$ companion and the black
hole companions in panel~(a). Here the double neutron star system (DNS)
orbital parameters
closely resemble the most relativistic binary pulsar system
currently known, the Double Pulsar, PSRs J0737$-$3039A/B, where the
l.o.s. acceleration tops out at $\sim 260$\,m\,${\rm s}^{-2}$. While
compact PSR-SBH systems of the same orbital period have maximum
acceleration values well above this ($\gtrsim$~600\,m\,${\rm s}^{-2}$)
the most extreme example is shown in the eccentric system where,
with the appropriate orientation of the major axis of the orbit to
the l.o.s. the acceleration can reach nearly 3000\,m\,${\rm s}^{-2}$
(see Fig.~\ref{fig:psr_bh_accels}, panel~c).

Secondly, in compact PSR-SBH systems (panel~a) the acceleration
derivatives (jerk) are significantly increased in comparison to DNS
of the same orbital period. Considering the first 0.25\,h of the
orbits in the systems with $\omega=0$, we find $|j|\sim
0.2,~0.6,~{\rm and}~0.9$\,m\,${\rm s}^{-3}$ for companion masses of
1.3, 10, and 30\,$M_{\odot}$ respectively. As we will show in
Section~5.3. despite acceleration searches performed on time series
that have a length where $T_{\rm obs} \lesssim P_{\rm b}/10$, these
higher order effects can still degrade the effectiveness of the
acceleration search method. It is also worth noting that pulsars in
eccentric systems, like those displayed in panel~(b) and (c), spend
the majority of the time in a state of low and constant orbital
acceleration. However, at periastron the acceleration and jerk can
reach values that might not be possible to correct for in real time
processing due to computational restrictions (see
Section~\ref{sssec:comp_scale}).

\subsubsection{Computational scaling}\label{sssec:comp_scale}
The dominant fraction of computing time in pulsar search processing
is typically spent at the acceleration search stage, where the
entire search process is essentially repeated a large number of
times for a number of trial accelerations. As such, consideration of
the level of computation required to find PSR-SBH with next
generation telescopes is needed to test the feasibility. Here we
limit our discussion to time-domain acceleration searches where the
approximate number of computational operations for acceleration
searches, $C_{\rm a}$, is proportional to the product of number of
trial accelerations $n_{\rm
  a}$ and the standard Fast Fourier Transformation (FFT) operation
count for one-dimensional real valued data,
\begin{equation}\label{eq:fft_count}
  C_{\rm a} \propto n_{\rm a}\,2.5\,n_{\rm samp}\,\log_2(n_{\rm samp})
\end{equation}
where $n_{\rm a}=\Delta a/\delta a$ and $n_{\rm samp}=T_{\rm
  obs}/\tau_{\rm samp}$; here $\Delta a$ is the total range of trial
accelerations searched, $\delta a$ is the step size in acceleration
over this range and $\tau_{\rm samp}$ is data sampling interval.
Estimating the exact number of computational operations in an
acceleration search is difficult since a pulsar search consists of not
just simple processes like FFTs, but many other operations, such as
spectral normalisation, harmonic summation, spectral interpolation and
the memory access and any associated input/output operations
\citep[for details of pulsar search processes,
  see][]{lk05}. Estimating the computational run time is even more
problematic and strongly depends on the hardware being used.
Representative tests typically only come from code benchmarking.
However, in the following we can derive the relevant computational
operation scaling for time domain acceleration searches; additional
overheads would scale in a similar manner.

Following \cite{clf+00} and choosing a step-size in acceleration
given by $\delta a = Pc/T_{\rm obs}^2$ ($P$ is the pulsar spin
period) we can write,
\begin{equation}\label{eq:ops}
C_{\rm a} \propto 2.5 \, \frac{\Delta a \, T_{\rm obs}^3}{P \, c \,
\tau_{\rm samp}}\,\log_2\left({\frac{T_{\rm obs}}{\tau_{\rm
samp}}}\right).
\end{equation}
Typically $P$ is set to the minimum spin period pulsar likely to be
detected (e.g. $\sim\,1\,{\rm ms}$) however to perform an optimal
search, one should consider not just the fundamental spin period but
the highest spin frequency harmonic that might be detected,
e.g. at $P/8$. The best solution is to consider the highest spin frequency
that can be detected, which is given by the Nyquist frequency, $f_{\rm
  Nyq}=1/2\tau_{\rm samp}$, giving a minimum spin period of
$2\tau_{\rm samp}$. Substitution for $P$ in Eq.~(\ref{eq:ops}) gives,
\begin{equation}\label{eq:opt_ops}
C_{\rm a} \propto 1.25 \, \frac{\Delta a \, T_{\rm obs}^3}{\, c \,
\tau_{\rm samp}^2}\,\log_2\left(\frac{T_{\rm obs}}{\tau_{\rm
samp}}\right).
\end{equation}
These relations are useful for considering the computational cost of
acceleration searches, and of particular importance for realtime
searches. For instance, it can be clearly seen that the integration
time has the biggest impact on the level of computation required ($C_{\rm a}
\propto T_{\rm obs}^3$); an important reason to keep $T_{\rm obs}$
as short as possible. In searches for PSR-SBH, $\Delta a$ should
encompass the maximum l.o.s. accelerations that might be observable
in these systems, like those displayed in
Fig.~\ref{fig:psr_bh_accels}: $\Delta a \approx \pm 1000\,{\rm m\,s}^{-2}$.
$\Delta a$ can in principle be reduced by a factor
${\rm sin}(i)$, where $i$ is the orbital inclination with respect to
the observer. For example a median inclination of $60^{\circ}$ could
be assumed, however, ensuring that acceleration searches can recover
the signal from PSR-SBH systems viewed edge-on is the ``safest''
option. Other reductions in the level of computation required can be
achieved by assuming larger values for $P$ or accounting for the
increasing values of $\tau_{\rm samp}$ typically implemented when
large dispersion measure (DM) trials are executed.

From Eq.~(\ref{eq:ops}) and following the SKA1-mid survey parameters
outlined in Table~24 of \cite{dtm+13} it can be shown that an
acceleration search for PSR-SBH systems, like those displayed in
Fig.~\ref{fig:psr_bh_accels} (i.e. where $\Delta a \approx \pm
1000\,{\rm m\,s}^{-2}$), would require at least an order of magnitude
more computational capacity than currently planned. An approximately
equivalent level of computation can be achieved either by adjusting
the acceleration step size to assume a minimum spin period of 20\,ms
\citep[c.f. 2\,ms
  in][]{dtm+13}, increasing the sampling interval by a factor of nine,
or by reducing the survey integration time from 600~s to
278~s. Alternative derivations of acceleration step-sizes can also
lead to reduced levels of computation \citep[see e.g.][]{ekl+13},
however, tests have shown that step-sizes which account for the
highest detectable spin frequency harmonic (as described above) have
better performance.

We also note that equivalent frequency domain acceleration searches
have the potential to probe the large values of acceleration that
might be observable in PSR-SBH, or indeed much higher values, for the
same or less computational cost \citep{rem02}. Frequency domain
methods offer this prospect because the acceleration values searched
by such algorimths are proportional to the spin period: $a=n_{\rm
  drift}cP/T_{\rm obs}^2$, here $a$ is the acceleration and $n_{\rm drift}$ is
the number of spectral bins drifted by the signal \citep[see
e.g.][]{hes07}. For example, for a spin period of 0.5~s and an
observation time of 600~s, a spectral bin drift of $n_{\rm
  drift}\sim2.4$ corresponds to an acceleration of
$1000\,{\rm m\,s}^{-2}$. Such values of $n_{\rm drift}$ are readily
probed with current computational hardware and
software\footnote{http://www.cv.nrao.edu/$\sim$sransom/presto/}
\citep[see e.g.][]{lrf+11}.

Initial surveys to be performed with FAST will include drift scan
surveys at lower frequencies \citep[400\,MHz,][]{yln13}. Because of
the extremely short integration time ($\sim 40$\,s), these surveys
will be sensitive to potential nearby PSR-SBH systems. Such short
integration lengths will also enable advanced binary searches
including jerk, or orbital parameters, to be done.

\subsection{Searches of simulated PSR-SBH}\label{ssec:sims}

We can investigate the effectiveness of acceleration searches in
recovering the pulsar signal in PSR-SBH. Establishing the
sensitivity of a pulsar acceleration search algorithm is a difficult
task since a number of parameters of both the binary system and the
observing system can affect the performance; examples include the
orbital period, orbital phase at which the observation was
performed, eccentricity of the orbit, companion mass, spin period
and observation length. To address this issue \cite{blw13} has
extended the work of \cite{jk91} to analytically define the expected
signal loss after acceleration searches or acceleration and jerk
searches for binary pulsar systems and observing systems of
arbitrary type. Here we have investigated the effectiveness of
acceleration search methods through simulations of the example
PSR-SBH systems described in Section~\ref{sssec:accn_char}.

\begin{figure*}
\centering
\includegraphics[scale=0.485]{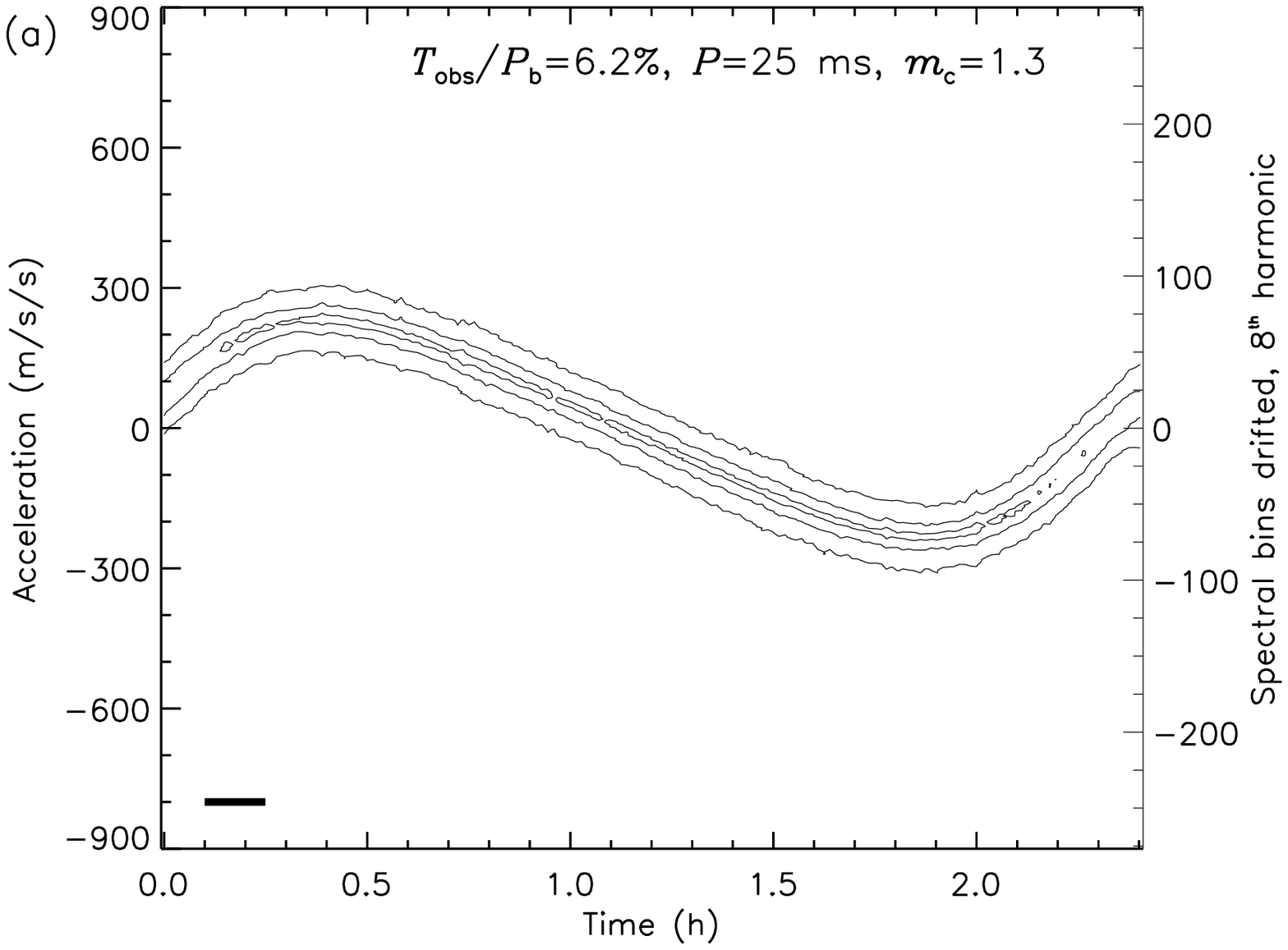}
\includegraphics[scale=0.485]{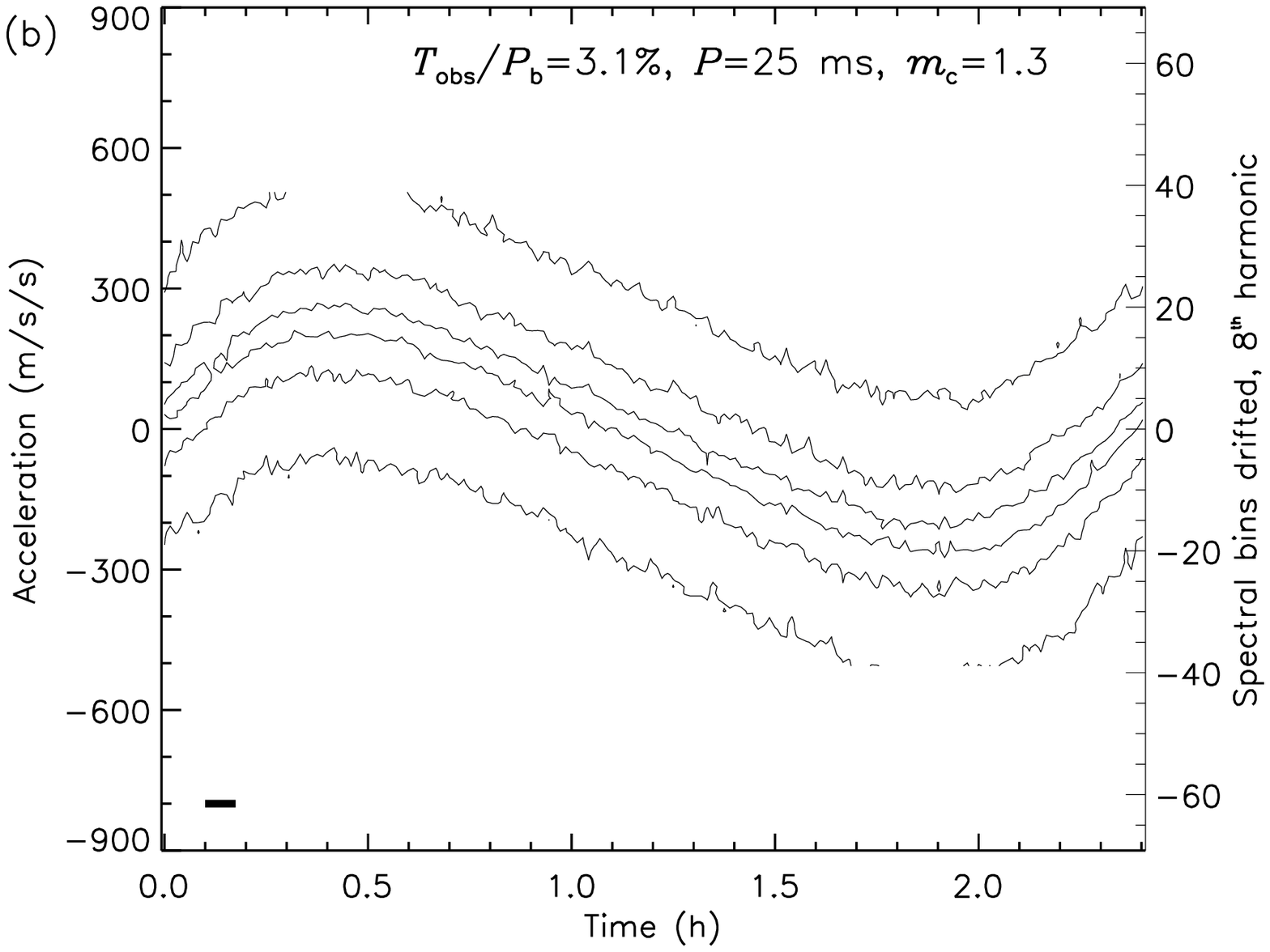}
\includegraphics[scale=0.485]{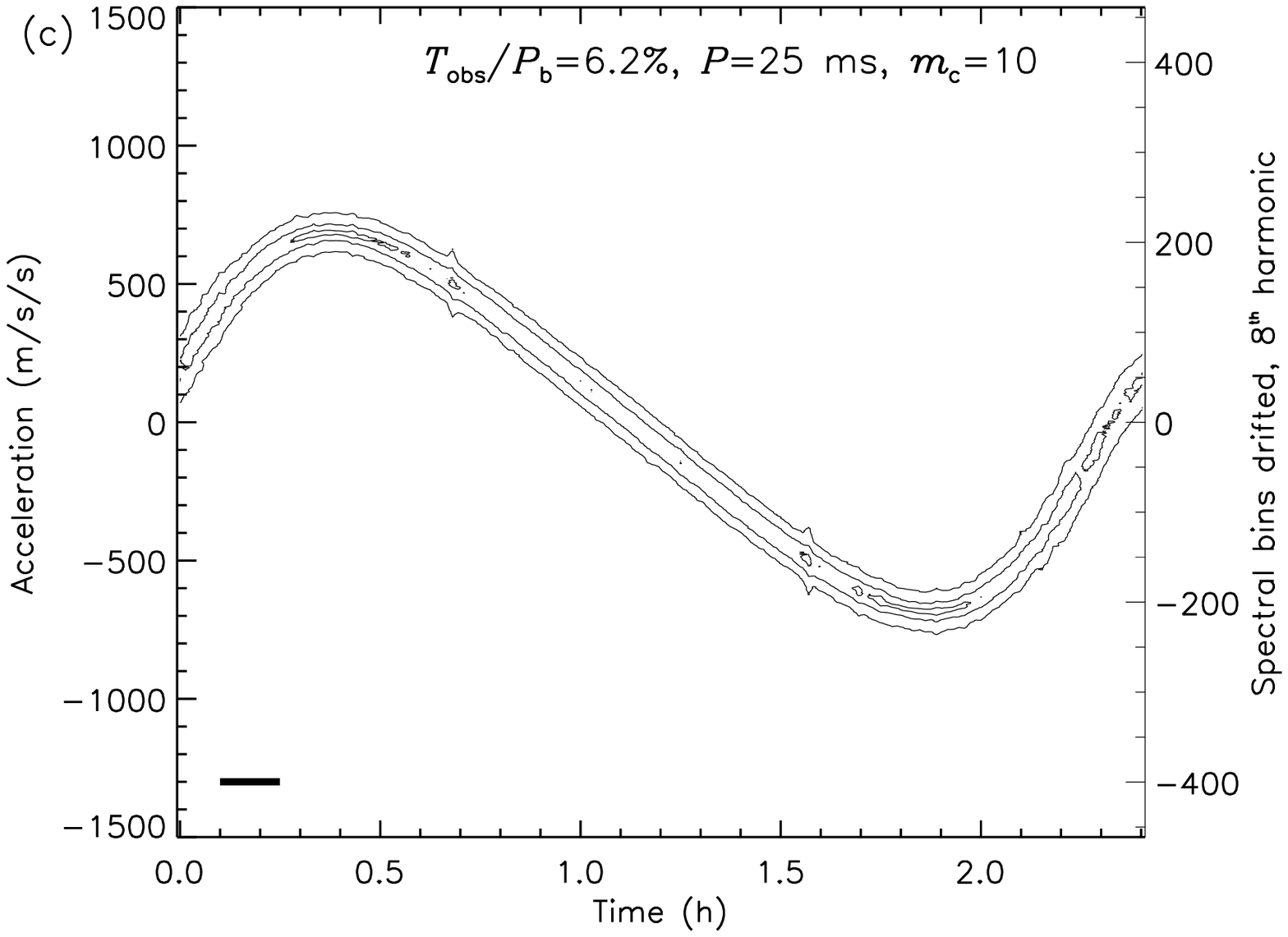}
\includegraphics[scale=0.485]{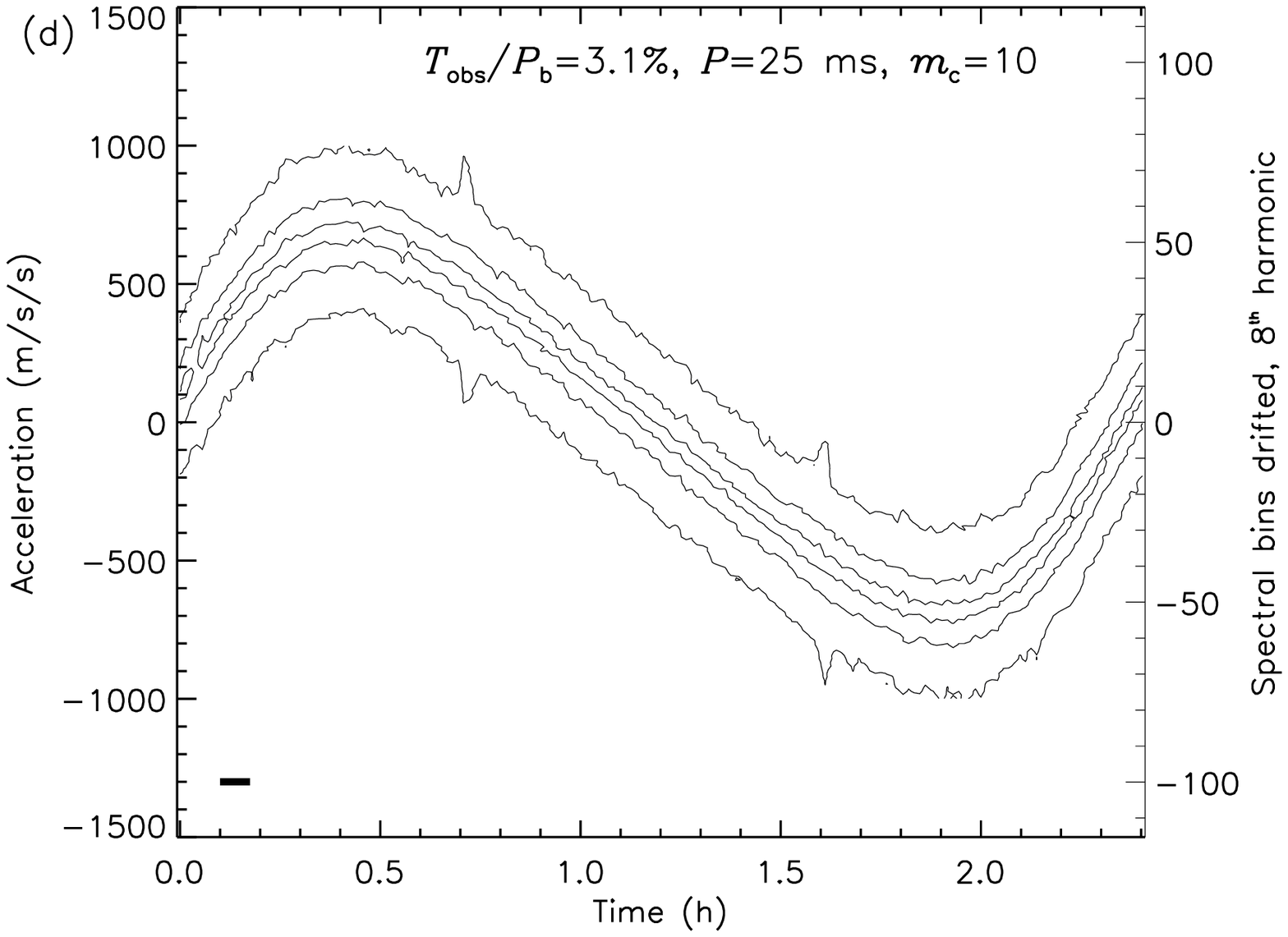}
\includegraphics[scale=0.485]{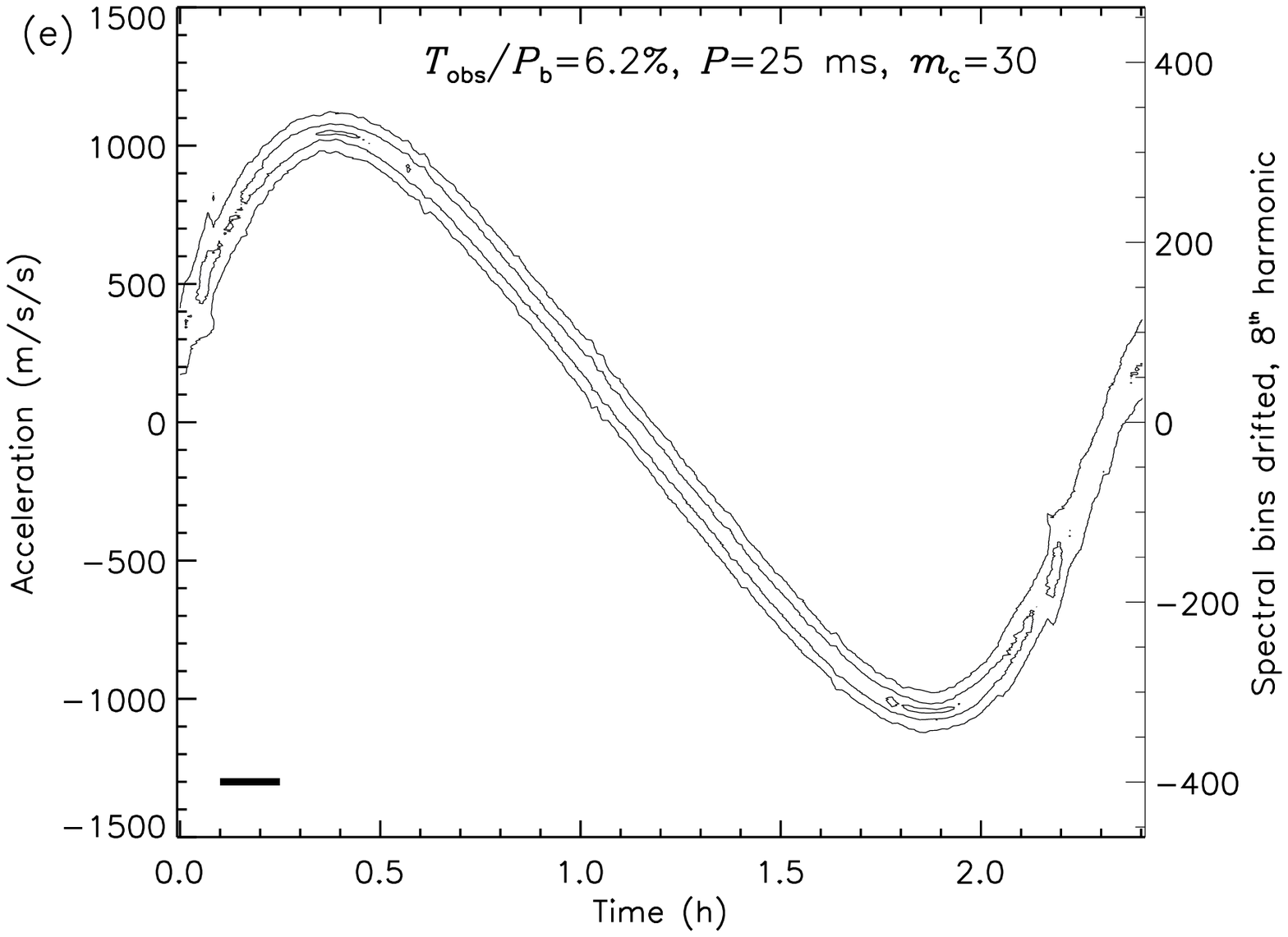}
\includegraphics[scale=0.485]{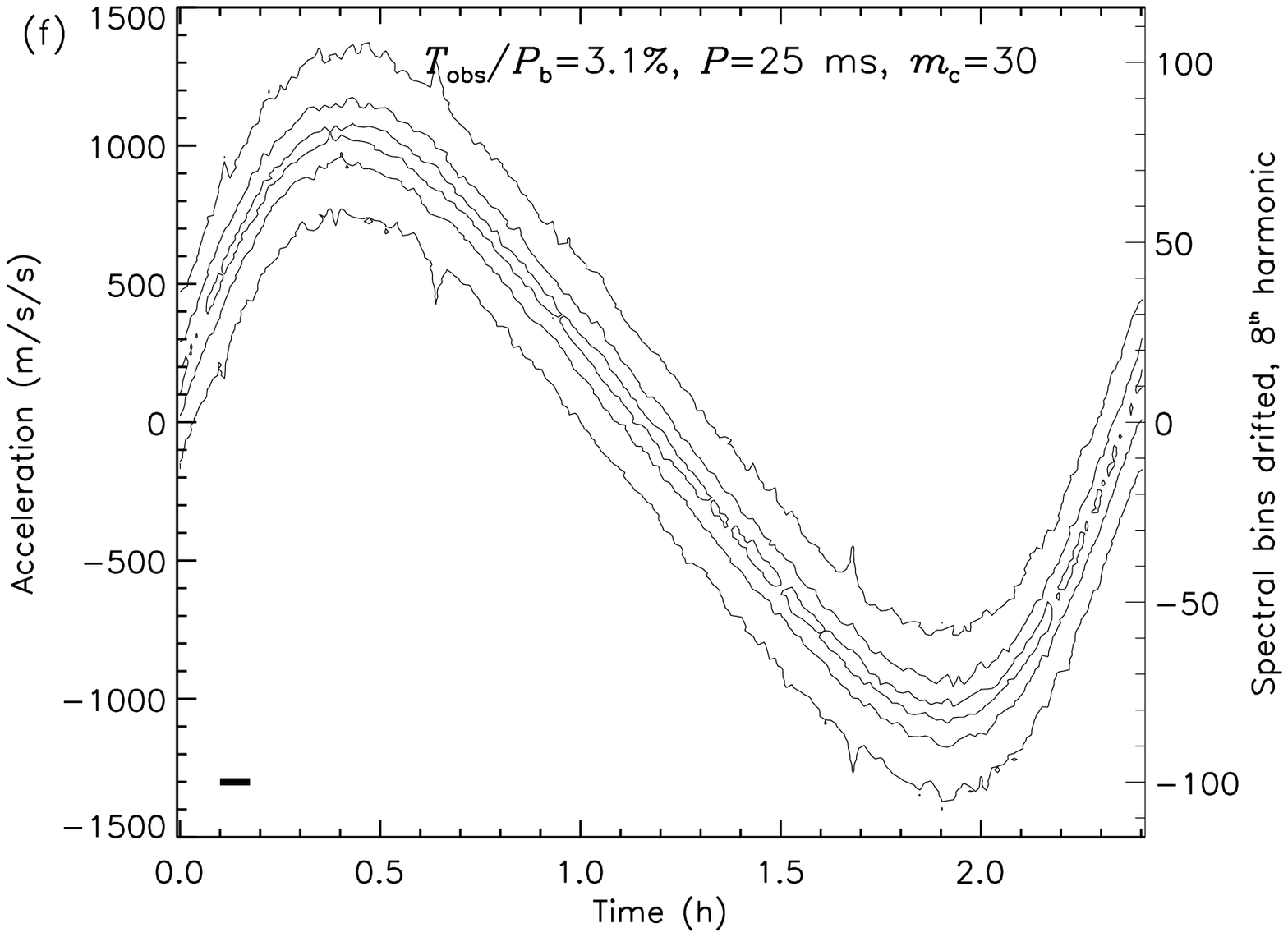}
\caption{Contours of the recovered spectral power achieved in
  acceleration searches, incremented in orbital phase, of simulated
  binary pulsar systems with orbital periods of $P_{\rm b}=2.4$\,h,
  $e=0.1$ and pulsar spin periods of 25\,ms. Results are plotted for
  pulsar companion masses of 1.3\,$M_{\odot}$ (Panels~a and b - DNS),
  10\,$M_{\odot}$ (Panels~c and d - PSR-SBH) and 30\,$M_{\odot}$
  (Panels~e and f - PSR-SBH), and integration times that cover $6.2$
  and $3.1$ per cent of the $2.4$\,h orbital period (panels~a, c, e
  and panels~b, d, f, respectively). Contours mark 30, 60, and 90
  percent recovery levels of the pulsar signal from an acceleration
  search.  The thick horizontal black bars indicate the length of
  integration used in the acceleration analysis. The starting point of
  each acceleration search was incremented in orbital phase by
  50~s. In panel b. 30 per cent recovery contours at
    extrema in acceleration are not visible due to the finite range of
    accelerations searched
    ($\pm\,506\,{\rm m}\,{\rm s^{-2}}$).\label{fig:S/Nrec-e0.1}}
\end{figure*}

In Fig.~\ref{fig:S/Nrec-e0.1}, we plot the acceleration value and
percentage of spectral power that can be recovered in constant
acceleration searches performed at various orbital phases, and with
two representative integration times. Contours mark the 30\%, 60\%
and 90\% recovery levels, where the percentage is that of the
spectral power collected in a search of an equivalent un-accelerated
solitary pulsar. Here we consider compact systems of low
eccentricity ($e=0.1$), where there is more chance the pulsar has
undergone a period of recycling. A spin period of 25~ms and a 4\%
pulse width have been assumed. In all cases the starting point of
the acceleration searches are incremented in time across the orbit
by 50\,s and an acceleration step-size, $\delta a$, that accounts
for the eighth pulsar harmonic has been chosen. On the right-hand
y-axis we give the number of spectral bins drifted ($n_{\rm drift}$)
by this eighth harmonic of the spin frequency; a number relevant to
acceleration searches performed in the frequency domain
\citep{rem02}. Both the simulation of data and the search are
conducted with an updated version of the \textsc{sigproc} software
package\footnote{http://sigproc.sourceforge.net/}. Panel~(a) shows
that for an observing time of 6.2\% of the orbital period
($\approx537$\,s) the search succeeds in recovering over 90\% of the
power at most orbital phases when the companion is of solar-mass.
However, for PSR-SBH systems observed with this integration time
(panels~c and e) the same level of recovery can only be achieved at
orbital phases where the variation of acceleration is minimised i.e.
at the peaks and troughs. When the observing time is halved to
$\approx268$\,s, 90\% signal recovery is achievable for observations
starting at most orbital phases and for companion masses up to
30\,$M_{\odot}$. Importantly, in terms of limiting telescope
sensitivity, the 60\% contours in the longer 537\,s integrations,
displayed in panels~(a), (c) and (e), are roughly equivalent to the
90\% contours in the shorter 268\,s integrations (panels~b, d and
f). Direct comparison of these contours shows that sensitivity
across the orbit is marginally worse in the longer integrations,
however, over eight times more computations are required to reach
this identical sensitivity level.

\begin{figure}\color{white}[]
\centering \color{black}
\includegraphics[scale=0.485]{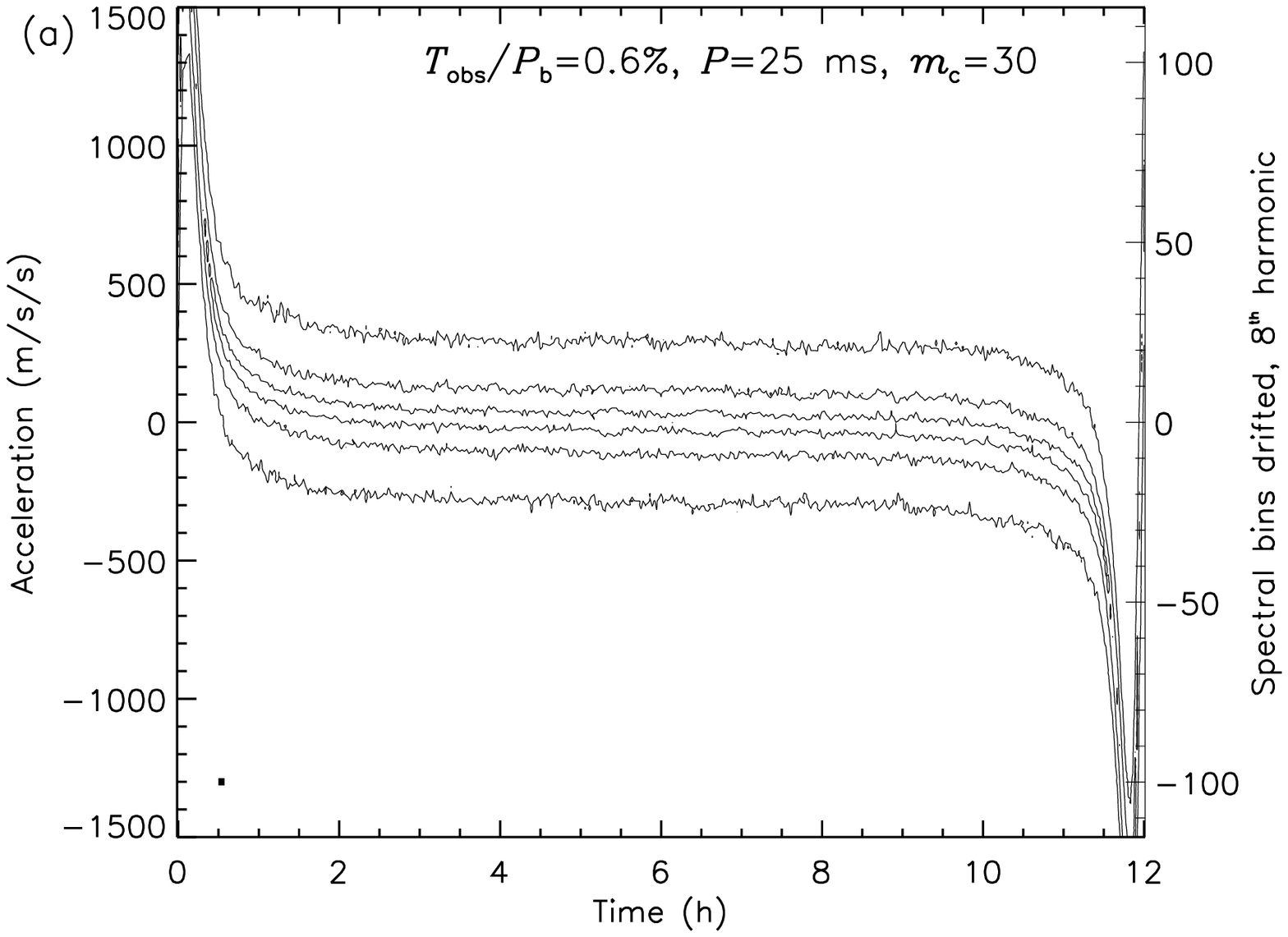}
\includegraphics[scale=0.485]{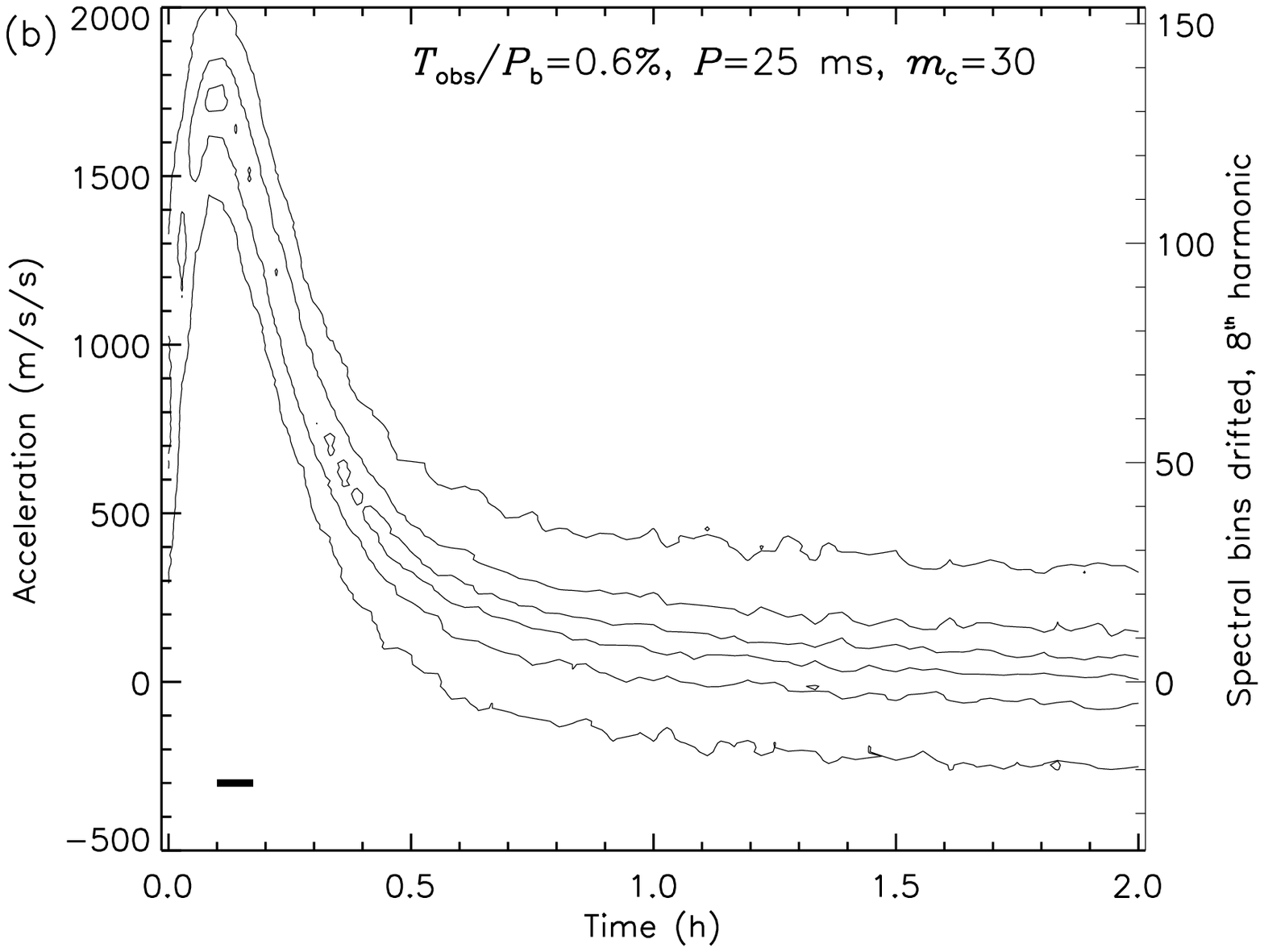}
\includegraphics[scale=0.485]{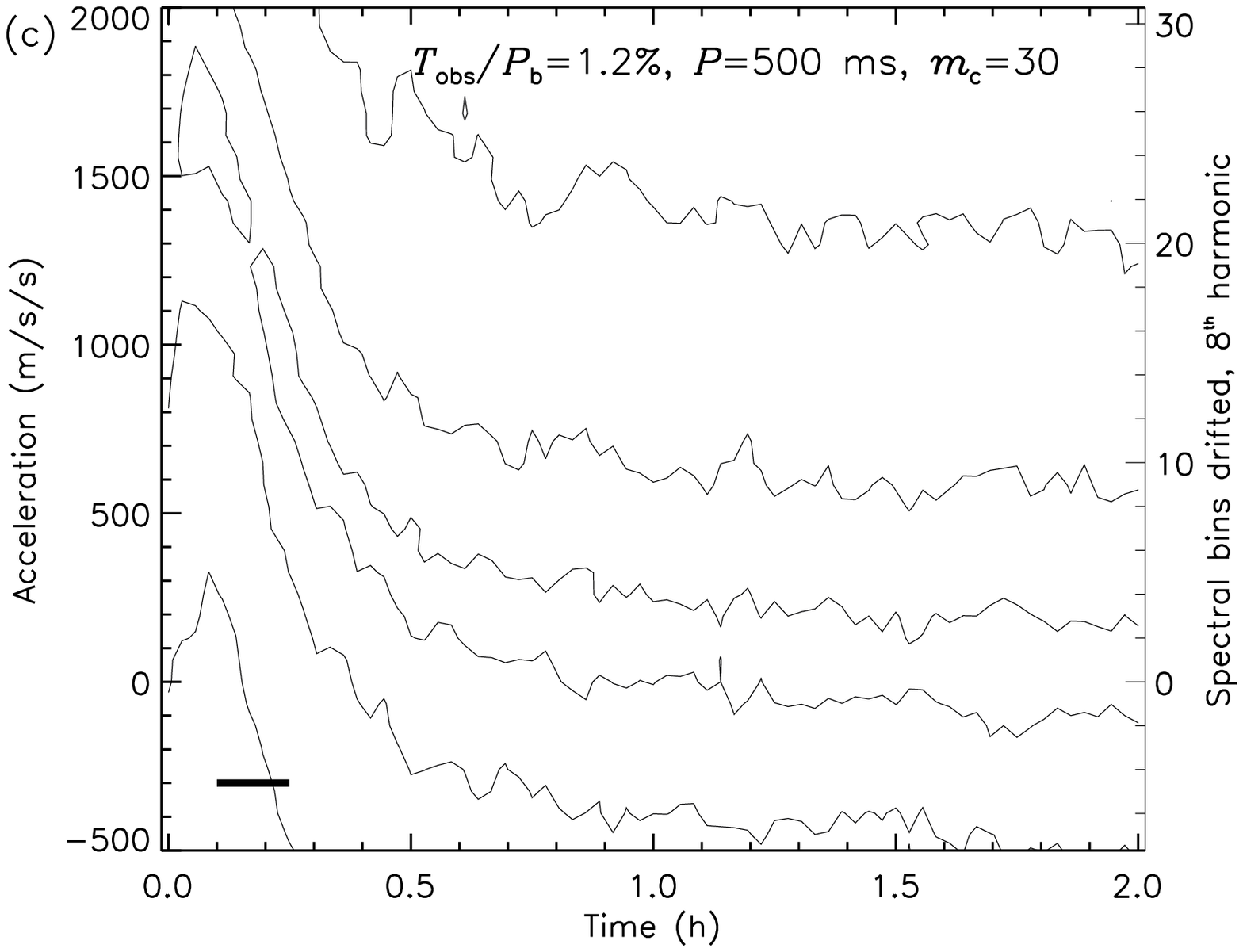}
\caption{Contours of the recovered spectral power achieved in
acceleration searches, incremented in orbital phase (by 100\,s), of a
simulated PSR-SBH system with an orbital period of $P_{\rm
b}=12.0$\,h and $e=0.8$. Panels~(b) and (c) show a zoom in of the
first 0.25\,h of the full orbit displayed in panel (a). In panel~(c)
the pulsar spin period is increased from 25\,ms (panels~a and b) to
0.5\,s. The representation of lines is the same as in
Fig.~\ref{fig:S/Nrec-e0.1}. \label{fig:S/Nrec-e0.8}}
\end{figure}

In Fig.~\ref{fig:S/Nrec-e0.8} the results of acceleration searches
of a simulated PSR-SBH system ($m_\bullet = 30$) with a wide
($P_{\rm b}=12.0$\,h) and eccentric ($e=0.8$) orbit is given. Both
slow pulsars ($P=0.5$\,s) and recycled ($P=25$\,ms) have been
investigated. Fig.~\ref{fig:S/Nrec-e0.8}, panel~(a) where $T_{\rm
  obs}\approx268$\,s, shows the recycled pulsar can be detected with
almost full sensitivity at relatively low accelerations ($|a|
\lesssim 100$\,m\,${\rm s}^{-2}$) throughout the majority ($\gtrsim
90$\%) of the orbit. However, close to periastron, as displayed in
panel~(b), acceleration searches, even of this reduced integration
time that works well in the low eccentricity systems
(Fig.~\ref{fig:S/Nrec-e0.1}, panel~f), is not short enough to
recover the pulsar signal. This indicates higher order corrections
are necessary for detection near periastron. For a longer spin
period pulsar ($P = 0.5$\,s) in the same system, displayed in
panel~(c), 90\% signal recovery is achieved for observations
starting much closer to periastron, even with longer integration
times ($T_{\rm obs}\approx537$\,s).

These simulations have demonstrated that within the scope of
constant acceleration searches for PSR-SBH, it is important to keep
a comparatively short observing time. Our results suggest that
integration times of the order 500\,s \citep[e.g.][]{sks+09,dtm+13}
to be performed with the SKA and SKA1 might be too long to guarantee full
instrumental sensitivity to the extreme PSR-SBH systems described
here; unless the pulsar has a longer spin period (e.g.\ 0.5\,s). Based
on probability arguments, high eccentricity PSR-SBH, containing
pulsars of any spin period, are easier to detect as the pulsar
spends the majority of time with low and constant l.o.s.
acceleration; such systems would also not be strongly selected
against with longer integrations. Nevertheless, our simulations
indicate that integration times of the order of $\sim 300$\,s seem to
be the maximum length with which most possible PSR-SBH systems can
be detected using constant acceleration search algorithms.

The great improvement in instrumental gain by the next generation of
radio telescopes would easily compensate for the loss of sensitivity
caused by shortening the integration time. For instance, the
currently on-going deep Galactic plane section of the High Time
Resolution Universe Pulsar Survey (HTRU-Deep) has an integration
time of 4300\,s with the Parkes Radio Telescope \citep[e.g.][ Ng et
al. in prep.]{kjv+10}. Assuming a telescope gain of $\sim
100$\,K/Jy, an SKA survey with 300\,s integration time would still
enable a factor of 20 to 30 increase in sensitivity. Following the
above comparison to HTRU-Deep we find a factor of two improvement in
sensitivity with a FAST drift scan survey.

%%%%%%%%%%%%%%%%%%%%%%%%%%%%%%%%%%%%%%%%%%%%%%%%%%%%%%%%%%%%%%%%%%%%%%%%%%%%%%%%

\section{Conclusions}
\label{sec:conclu}

In this paper, we investigate the achievable gravity tests by
observing a PSR-SBH binary system, in particular with the next
generation of radio telescopes. For our studies, we have used the
sensitivity of the FAST and SKA as representatives for the future
telescopes. The investigations are based on simulated pulsar timing
data with consistent timing models. It has been shown that with
three to five years pulsar timing observations we can expect to
measure the masses of the pulsar and the black hole with a high
precision (e.g., $0.001\,\%-1\%$), especially when the pulsar is a
MSP and the observations are conducted with the next generation of
radio telescopes. The black hole spin can be measurable on
timescales of five to ten years, with high precision ($\sim1\%$) on
the same optimal scenario. Those measurements will lead to a test of
GR's cosmic censorship conjecture. The quadrupole of the black hole
is measurable only when the pulsar is fully recycled and the system
is of extreme configuration, in terms of orbit compactness or mass
of the black hole. In addition, we showed that a PSR-SBH system
would enable a few orders of magnitude improvement in constraining
alternative gravity theories which predicts practically identical
black holes to GR. This is demonstrated with the help of a specific
class of scalar-tensor theories of gravity.

Finally, we have investigated some of the requirements in searches
for PSR-SBH. The large instantaneous sensitivity of next generation
of radio telescopes should allow reduced pulsar survey integration
times, thereby enabling the potentially extreme orbital parameter
space associated with PSR-SBH to be searched more effectively.

%%%%%%%%%%%%%%%%%%%%%%%%%%%%%%%%%%%%%%%%%%%%%%%%%%%%%%%%%%%%%%%%%%%%%%%%%%%%%%%%

\section*{Acknowledgements}

We thank P.~C.~C.~Freire and S.~M.~Ransom for useful discussions and
reading the manuscript. We are grateful to J.~Lattimer for providing
the tables for equations-of-state MPA1. The authors also wish to
thank the anonymous referee for valuable comments. K.~Liu is
supported by the ERC Advanced Grant ``LEAP", Grant Agreement Number
227947 (PI M.~Kramer).

%%%%%%%%%%%%%%%%%%%%%%%%%%%%%%%%%%%%%%%%%%%%%%%%%%%%%%%%%%%%%%%%%%%%%%%%%%%%%%%%

\bibliographystyle{mnras}
\bibliography{journals_apj,psrrefs,modrefs,crossrefs}

\end{document}